\documentclass[acmsmall]{acmart}
\usepackage{stfloats}
\usepackage{makecell}
\usepackage{multirow}
\usepackage{graphicx}
\usepackage{subfigure}
\usepackage{pifont}
\usepackage{listings}
\usepackage{threeparttable}
\definecolor{revise}{RGB}{0,0,0}
\definecolor{REVISE}{RGB}{0,0,0}
\definecolor{minorrevise}{RGB}{0,0,0}
\definecolor{MINORREVISE}{RGB}{0,0,0}
\definecolor{toconfirm}{RGB}{0,0,0}
\definecolor{TOCONFIRM}{RGB}{0,0,0}

\AtBeginDocument{%
  }

\setcopyright{rightsretained}
\copyrightyear{2024}
\acmYear{2024}
\acmDOI{XXXXXXX.XXXXXXX}

\begin{document}
\title[Prompt2Task]{\textcolor{minorrevise}{Prompt2Task: Automating UI Tasks on Smartphones from Textual Prompts}}

\author{Tian Huang}
\email{ht20@mails.tsinghua.edu.cn}
\orcid{0009-0007-8639-5929}
\affiliation{%
  \institution{Department of Computer Science and Technology, Tsinghua University}
  \city{Beijing}
  \country{China}
}

\author{Chun Yu}
\orcid{0000-0003-2591-7993}
\authornote{Corresponding author.}
\affiliation{%
  \institution{Department of Computer Science and Technology, Tsinghua University}
  \city{Beijing}
  \country{China}
}
\email{chunyu@tsinghua.edu.cn}

\author{Weinan Shi}
\orcid{0000-0002-1351-9034}
\affiliation{%
  \institution{Department of Computer Science and Technology, Tsinghua University}
  \city{Beijing}
  \country{China}
}
\email{swn@mail.tsinghua.edu.cn}

\author{Zijian Peng}
\orcid{0009-0005-6971-0064}
\affiliation{%
  \institution{Department of Computer Science and Technology, Tsinghua University}
  \city{Beijing}
  \country{China}
}
\email{pzj21@mails.tsinghua.edu.cn}

\author{David Yang}
\orcid{0000-0001-8260-3830}
\affiliation{%
  \institution{Department of Computer Science and Technology, Tsinghua University}
  \city{Beijing}
  \country{China}}
  \email{ydw22@mails.tsinghua.edu.cn}

\author{Weiqi Sun}
\orcid{0009-0009-0957-4864}
\affiliation{%
  \institution{Department of Public Administration, Sichuan University}
  \city{Sichuan}
  \country{China}}
  \email{weiqiviki999@gmail.com}

\author{Yuanchun Shi}
\orcid{0000-0003-2273-6927}
\affiliation{%
  \institution{Department of Computer Science and Technology, Tsinghua University, Qinghai University}
  \city{Beijing}
  \country{China}
}
\email{shiyc@tsinghua.edu.cn}

\renewcommand{\shortauthors}{Huang et al.}




\begin{abstract}

\textcolor{minorrevise}{UI task automation enables efficient task execution by simulating human interactions with graphical user interfaces (GUIs), without modifying the existing application code.} However, its broader adoption is constrained by the need for expertise in both scripting languages and workflow design. To address this challenge, we present Prompt2Task, a system designed to comprehend various task-related textual prompts (e.g., goals, procedures), thereby generating and performing the corresponding automation tasks. Prompt2Task incorporates a suite of intelligent agents that mimic human cognitive functions, specializing in interpreting user intent, managing external information for task generation, and executing operations on smartphones. The agents can learn from user feedback and continuously improve their performance based on the accumulated knowledge. Experimental results indicated a performance jump from a 22.28\% success rate in the baseline to 95.24\% with Prompt2Task, requiring an average of 0.69 user interventions for each new task. Prompt2Task presents promising applications in fields such as tutorial creation, smart assistance, and customer service.


\end{abstract}




\maketitle

\section{Introduction}

\textcolor{minorrevise}{UI task automation has emerged as a pivotal technology, significantly boosting productivity and operational efficiency \cite{10.1145/3025453.3025483}.} Its ability to simulate human interactions on the graphical user interface (GUI) allows for broad compatibility with third-party applications, without necessitating modifications to existing code \cite{deka_rico_2017, jovanovic2018robotic}. However, task automation primarily relies on predefined operation sequences \cite{RIBEIRO202151, herm2023framework}. Customizing these sequences often requires specialized skills in scripting languages and workflow design, which is a significant barrier to adoption, particularly for users without a technical background \cite{10.1007/978-3-319-66963-2_7}. Additionally, the fixed operation sequences have difficulty adapting to the frequently updated GUI, making them susceptible to becoming quickly outdated.



\begin{figure}[b]
\centering
    \includegraphics[width=\textwidth]{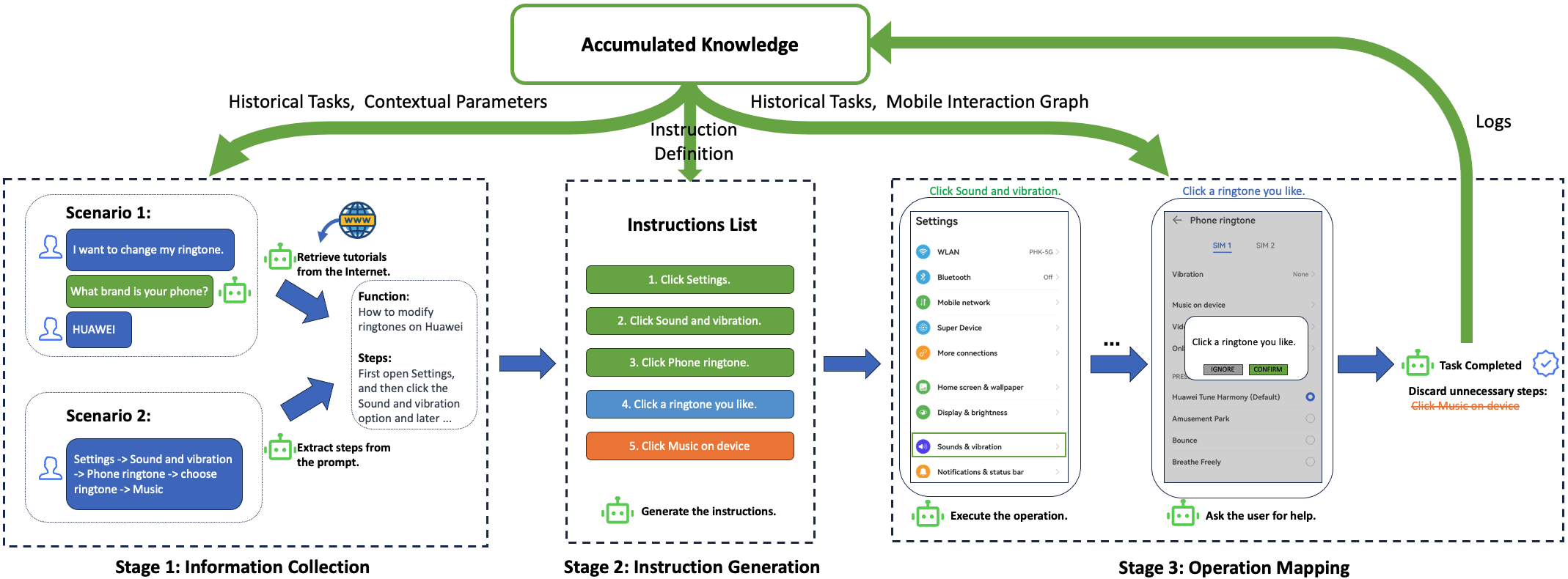}
  \caption{
  The workflow of Prompt2Task. \textcolor{revise}{In the first stage, users input task-related textual prompts, and the agent collects relevant textual information to plan the task. In the second stage, the agent converts the text into formally defined instructions. In the third stage, the agent maps these instructions to operations, ultimately executing the automation task on the smartphone. Users can intervene in the agents' decisions, especially when help is requested.} Blue arrows guide the process from prompts to task automation, while green arrows indicate the knowledge accumulation through user interactions, thereby enhancing the future performance of the agents.}
  \label{fig:title}
\end{figure}

\begin{sloppypar}
To address this gap, we introduce Prompt2Task, which employs a text-prompt-driven task automation approach implemented on smartphones. Prompt2Task is capable of accepting unrestricted textual descriptions for UI tasks---from concrete operational instructions like ``open settings, locate WLAN, and...'' to high-level functional requests such as ``change the ringtone''. Prompt2Task then automatically translates these textual prompts into operation sequences on GUI, freeing users from the complexity of designing automation scripts themselves.
\end{sloppypar}

Prompt2Task serves as an intelligent navigation system on smartphones. It interprets user intent from textual descriptions and matches them to its internal ``map'' for route planning and decision-making, thereby automatically generating task flows. Throughout the process, Prompt2Task faces several challenges. \textcolor{revise}{Firstly, it needs to accommodate diverse and unclear textual inputs, requiring Prompt2Task to infer intentions from vague or incomplete instructions \cite{10.1145/371920.371974}. Secondly, the ``map'' may be incomplete, especially lacking detailed route information for less common tasks or newly added application functions \cite{islam2010mobile}. This issue is difficult to address even with state-of-the-art Large Language Models (LLMs), which may generate incorrect or misleading information in the GUI domain \cite{10.1145/3571730, wang_enabling_2023}. Thirdly, it is challenging to adapt to frequent updates in interfaces and applications; as they evolve, the routes Prompt2Task relies on may also require updates \cite{venkatesh_ugif_2022}. These challenges span across different stages of the process, requiring the application of various types of intelligence and knowledge to address them effectively.}



To meet these challenges, we employ a multi-agent framework \cite{panait2005cooperative} for Prompt2Task. The framework allocates specialized agents to sub-tasks within different stages of the process, with each agent emulating specific human cognitive functions. As illustrated in Figure \ref{fig:title}, Prompt2Task spans three stages---from \textbf{information collection} and \textbf{instruction generation} to \textbf{operation mapping}. Specifically, \textbf{information collection} involves gathering all necessary descriptions of functions and steps for the automation task. \textbf{Instruction generation} entails creating formalized instructions to guide operations, and \textbf{operation mapping} refers to translating text-based instructions into operations on smartphones to ultimately complete the task. 

Two core elements are critical for Prompt2Task's performance: a comprehensive knowledge base \cite{venkatesh_ugif_2022}, which is our smartphone ``map'', and stable decision-making \cite{zhong_helpviz_2021}, which ensures the agents' efficient and collaborative task execution. Therefore, we create an evolving knowledge base that expands as it accumulates data and user interactions. We introduce three key agents to meet the specific knowledge demands of UI task automation: the \emph{Analysis Agent}, which identifies user intents and actively fills in missing contextual parameters from user inputs; the \emph{Retrieval Agent}, which sources online tutorials to guide task automation; and the \emph{Grounding Agent}, which conducts in-depth semantic analysis of mobile GUI for better alignment with textual inputs and supports unplanned operations upon updates or errors. To enhance decision-making, the \emph{Assessment Agent} scrutinizes decisions made by other agents, serving as an internal audit mechanism to improve accuracy and reliability. Beyond internal audit, the agents are seamlessly integrated into a user interface, allowing for user oversee and intervention to enhance decisions. \textcolor{revise}{User intervention is optional and involves chatting, selecting, editing, or demonstrating to modify and confirm the agents' decisions.} User usage and intervention create a feedback loop for continuous knowledge accumulation and system improvement, thereby optimizing efficiency for future tasks.

To evaluate Prompt2Task, we conducted a performance evaluation with 2,500 textual prompts crowd-sourced. The results showed that Prompt2Task increased the task success rate from a baseline of 22.28\% to 95.24\%, requiring only an average of 0.69 user interventions for each new task. Moreover, it was proved that the accumulation of user usage and intervention improved system performance. Furthermore, a user study involving both skilled and unskilled smartphone users confirmed Prompt2Task's effectiveness in assistance and its low interaction cost. Due to the low-cost translation from text to automation tasks, Prompt2Task has promising applications in generating interactive tutorials, enhancing smart assistance, and improving online customer service.



\textcolor{revise}{Specifically, the contributions of this work are three-fold.
\begin{itemize}
    \item We propose Prompt2Task, a multi-agent system that accepts unrestricted textual expressions of task requirements and automates corresponding UI tasks. To improve the success rate, Prompt2Task incorporates several innovative strategies: retrieving online tutorials, integrating necessary user interventions, and continuously improving agent capabilities through the accumulation of knowledge from use.
    \item We propose an effective approach to minimize the cost of user interventions. Prompt2Task presents results stage-by-stage, allowing users to intuitively interact with the system through dialogues, selections, edits, and demonstrations. A dedicated agent employs GPT-4 to evaluate the confidence of predicted operations, requesting user assistance only when necessary.
    \item We create a Chinese dataset \footnote{The dataset is at \href{https://github.com/PromptRPA/Prompt2TaskDataset}{https://github.com/PromptRPA/Prompt2TaskDataset}.} that covers a wide range of UI tasks and collects diverse user descriptions of the tasks. The dataset includes 2,500 textual prompts, spanning 100 tasks across 10 application domains, with a total of 533 instructions and 543 operations.
\end{itemize}}

\section{Related Work}

Our work intersects with four research areas: prior UI task automation systems, the knowledge required for UI task automation, LLM-driven multi-agent frameworks, and incorporating user feedback for agent improvement.

\subsection{Prior UI Task Automation Systems}

Prior UI task automation systems on smartphones typically work by replaying predefined interaction flows on graphical user interfaces, thereby completing the automation tasks. Early systems like SUGILITE \cite{10.1145/3025453.3025483} and APPINITE \cite{li_appinite_2018} utilized programming by demonstration to enable users to create interaction flows. However, for complex or unfamiliar functions, creating flows is challenging, especially for beginners and the elderly, who may lack essential task execution skills \cite{chen_design_2017, 10.1145/3550321}. Subsequent developments introduce interactive, context-aware tutorials for GUI applications \cite{10.1145/3290605.3300527,10.1145/2047196.2047213, li_intelligently_2015, zhong_helpviz_2021, li_learning_2021}. However, these systems primarily rely on pre-configured flows by developers \cite{wang_evertutor_2014}. \textcolor{revise}{When accommodating new tasks or when interface updates cause the original flows outdated \cite{10.1145/3586183.3606824}, these systems require users to have specialized skills in script design \cite{9233206}.} 


Recent efforts have focused on translating natural language into operation sequences on smartphones. Notable contributions include \citeauthor{branavan_reinforcement_2009} \cite{branavan_reinforcement_2009} and \citeauthor{li_mapping_2020} \cite{li_mapping_2020}. However, they require detailed instructions for each step, necessitating that users have a thorough understanding of task execution. While there are systems that support general task requests, limitations exist, such as UGIF's reliance on a pre-collected tutorial dataset \cite{venkatesh_ugif_2022} and Voicify's dependence on app developers' implementation of deep links \cite{10.1145/3581998}, failing to meet a wide range of user needs. \textcolor{revise}{Additionally, they often function as ``black boxes'', which can autonomously complete tasks when functioning correctly but may also produce uncontrollable and erroneous results \cite{venkatesh_ugif_2022}.}





\subsection{Knowledge Required for UI Task Automation}

Successful automation of smartphone tasks demands a comprehensive understanding of diverse knowledge domains. For the input, tutorial datasets need to be sufficiently rich to support the diversity of user needs. Regarding the output, a nuanced grasp of mobile semantics is required to ensure robust execution on smartphones.

Addressing user input presents challenges in accuracy and scope. User inputs are often vague \cite{jansen_real_1998}, requiring supplementary contextual information for accurate recognition of user intent \cite{ 10.1145/3290605.3300439}. In terms of scope, existing UI trace datasets such as Rico \cite{deka_rico_2017}, UGIF-DataSet \cite{venkatesh_ugif_2022}, and PixelHelp \cite{li_mapping_2020} are pre-collected and \textcolor{revise}{they are insufficient to meet a wide array of dynamic user needs \cite{10.1145/3290605.3300473}.} AutoVCI \cite{pan_automatically_2022} and Kite \cite{10.1145/3210240.3210339} can only recognize user intents limited to predefined UI trace collections. When using search engines or AI tools like GPT for tutorial information, the results can be irrelevant, outdated \cite{venkatesh_ugif_2022}, fabricated \cite{shen2023chatgpt}, or low-quality \cite{burns_dataset_2022}.

The other focus is the understanding of mobile GUI semantics, which is crucial for widget identification and context comprehension. A wealth of research focuses on providing semantic text annotations for UI elements \cite{ li_widget_2020, he_actionbert_2021, chen_towards_2022, wang_enabling_2023}. Other research efforts include generating full-page text descriptions \cite{wang_screen2words_2021}, image-to-HTML conversion \cite{lee_pix2struct_2022}, and GUI embedding \cite{ijcai2021p235, li_screen2vec_2021}. \textcolor{revise}{Translating GUI images and hierarchical information into natural language descriptions while preserving key task-relevant information is crucial for \textbf{operation mapping}, as it enables textual instructions to align effectively with GUI operations.}




\subsection{LLM-Driven Multi-Agent Frameworks}
Large Language Models (LLMs) have been widely applied in decision-making tasks recently. For one thing, some studies focus on the reasoning capabilities of LLMs. ReAct \cite{yao2022react} investigates the generation of reasoning traces and actions, and ExpeL \cite{zhao2023expel} delves into the realm of autonomous experiential learning. For another, increasing works aimed at enhancing the LLMs' interactive capabilities with external environments. Toolformer \cite{schick2023toolformer} enables LLMs to interact with APIs, whereas \citeauthor{dipalo2023unified} \cite{dipalo2023unified} explores their role within reinforcement learning agents. ``Do As I Can, Not As I Say'' \cite{ahn2022i} introduces a grounding mechanism, ProgPrompt \cite{10161317} targets robotic task planning, and VisualGPT \cite{Chen_2022_CVPR} examines multi-modal adaptability. Moreover, several works focus on multi-agent cooperation driven by LLMs. The frameworks proposed by \citeauthor{hong2023metagpt} \cite{hong2023metagpt}, \citeauthor{du2023improving} \cite{du2023improving}, \citeauthor{liang2023encouraging} \cite{liang2023encouraging}, \citeauthor{wei2023multiparty} \cite{wei2023multiparty} and \citeauthor{zhang2023building} \cite{zhang2023building}, explore various aspects ranging from collective decision-making to divergent thinking. 

\textcolor{revise}{Prior work has shown the effectiveness of multi-agent frameworks in addressing complex tasks, which we have applied in designing Prompt2Task.}
 


\subsection{Incorporating User Feedback for Agent Improvement}


Interactive machine learning (IML) \cite{10.1145/604045.604056, 10.1145/3185517, amershi_power_2014} aims to achieve complementary benefits by integrating human perception and intelligence with the computational power of computers \cite{deng_integrating_2020}. The framework has seen extensive application in fields requiring annotation \cite{sperrle_viana_2019}, classification \cite{berg2019ilastik, de2020interactive}, and intricate data analysis \cite{10.1145/3334480.3382839}. \textcolor{revise}{In line with the principles of IML, Prompt2Task employs an ensemble of agents that evolve through user feedback. The key problem is how to incorporate minimal, low-cost user intervention to achieve the maximum improvement in system performance.}

\section{\textcolor{minorrevise}{Problem of UI Task Automation}}
\label{section:pipeline}

UI task automation is traditionally performed by mimicking predefined, rule-based operations on graphical user interfaces \cite{10.1145/3025453.3025483}. Recent advances in natural language processing and human-centric design have indicated a more dynamic, context-aware era in the task automation, guided by textual prompts. These textual inputs are not only easily formulated but also can offer valuable information, such as the task's goal, steps, related parameters, and specific applications. 

\textcolor{toconfirm}{In this section, we introduce Prompt2Task model to better understand the UI task automation process driven by text. Based on this model, we analyze the challenges within the task automation, which inform our system's architecture.}






\subsection{Prompt2Task Model}
\label{sec:rpadefinition}


    
    
    



\begin{figure}[htbp]
  \centering
  \includegraphics[width=\linewidth]{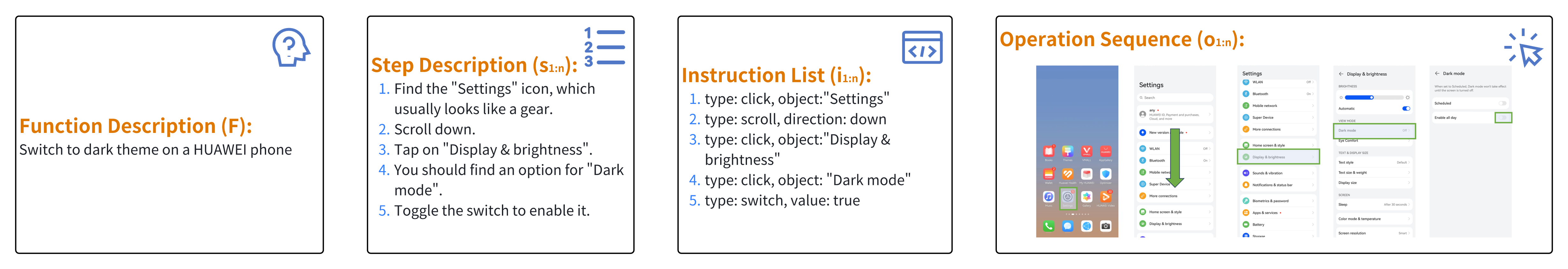}
  \caption{Components of a Prompt2Task model.}
  \label{fig:rpadefinition}
\end{figure}




\textcolor{minorrevise}{Prior task models were developed to provide formalized descriptions of tasks, typically in the context of user interface design. They are primarily used to describe complex system interactions and workflows, such as those involving concurrency and collaboration in high-complexity tasks \cite{limbourg2004comparing}. To address task automation on mobile GUI, we propose a simplified UI task model based on ConcurTaskTrees \cite{10.5555/647403.723688}, as shown in Figure \ref{fig:rpadefinition}. This model consists of the following core components:}

\begin{itemize}
    \item \textbf{Function Description ($F$)}: A piece of natural language text that represents the overarching objective the automation task aims to achieve.
    
    \item \textbf{Step Description ($s_{1:n}$)}: A list of textual items detailing individual steps within the automation process, each specified in natural language.
    
    \item \textbf{Instruction List ($i_{1:n}$)}: A list of formalized instructions, derived from natural language text, encoded in a machine-friendly standard language to direct the operation sequence.
    
    \item \textbf{Operation Sequence ($o_{1:n}$)}: An ordered set of operations required by the automation task, dynamically adjusted based on the textual prompts and the real-time operating environment.
\end{itemize}


\textcolor{minorrevise}{In the Prompt2Task model, $F$ and $s_{1:n}$ are used to express the task's goal and steps through natural language. While natural language lacks the rigor of prior task models such as the tree structure of ConcurTaskTrees \cite{10.5555/647403.723688} or the precise descriptions of concurrency, communication, and temporal order in LOTOS \cite{BOLOGNESI198725}, it provides sufficient flexibility to describe the core requirements of a task effectively. More importantly, it reduces the complexity for users, making it more accessible for non-experts. Additionally, $i_{1:n}$ and $o_{1:n}$ in the Prompt2Task model are represented as linear lists, since the execution of each automation task is typically carried out on a single device through simulated operations, making task execution inherently linear. Even when task planning involves branching structures, the actual execution remains linear. It does not require the complex structure in Hierarchical Task Analysis \cite{STANTON200655} or concurrency mechanisms in Event-Driven Process Chains \cite{scheer2005process}. By generating the operation sequence dynamically, the model is flexible enough to adapt to real-time changes. Therefore, the Prompt2Task model is well-suited to handle the dynamic and diverse requirements of UI task automation efficiently.}

\subsection{\textcolor{revise}{Challenges for UI Task Automation}}






\textcolor{revise}{In this paper, we focus on the process of inputting a textual prompt $P$ to generate the aforementioned Prompt2Task model and ultimately execute the automation task based on the model. The information contained in $P$ is insufficient to accurately generate the model components. Therefore, the process necessitates the integration of additional knowledge, such as application-specific details, user context, and current GUI state. In this process, we identify the following four challenges:}


\textcolor{revise}{\textbf{1. Input:} The input can be any text related to the task, ranging from formal manuals and tutorials to diverse user inputs. Users' expressions vary greatly, influenced by their differing levels of knowledge. Developers or skilled users can provide clear, standard instructions for specific operations, while novices or elderly users often struggle to articulate even basic functional requirements, sometimes only having a vague concept of the desired outcome \cite{10.1145/3550321}. Even skilled users might expect to automate tasks with minimal and simple inputs. Most current research \cite{pasupat2018mapping, li_mapping_2020, pan_automatically_2022} accepts explicit, step-by-step commands, such as ``click the confirm button'' or ``input Hello in the edit box'', but falls short in handling more complex commands like ``activate developer mode'', and are particularly challenged by ambiguous requests like ``change the appearance''.} 

\begin{figure}[t]
  \centering
  \includegraphics[width=\linewidth]{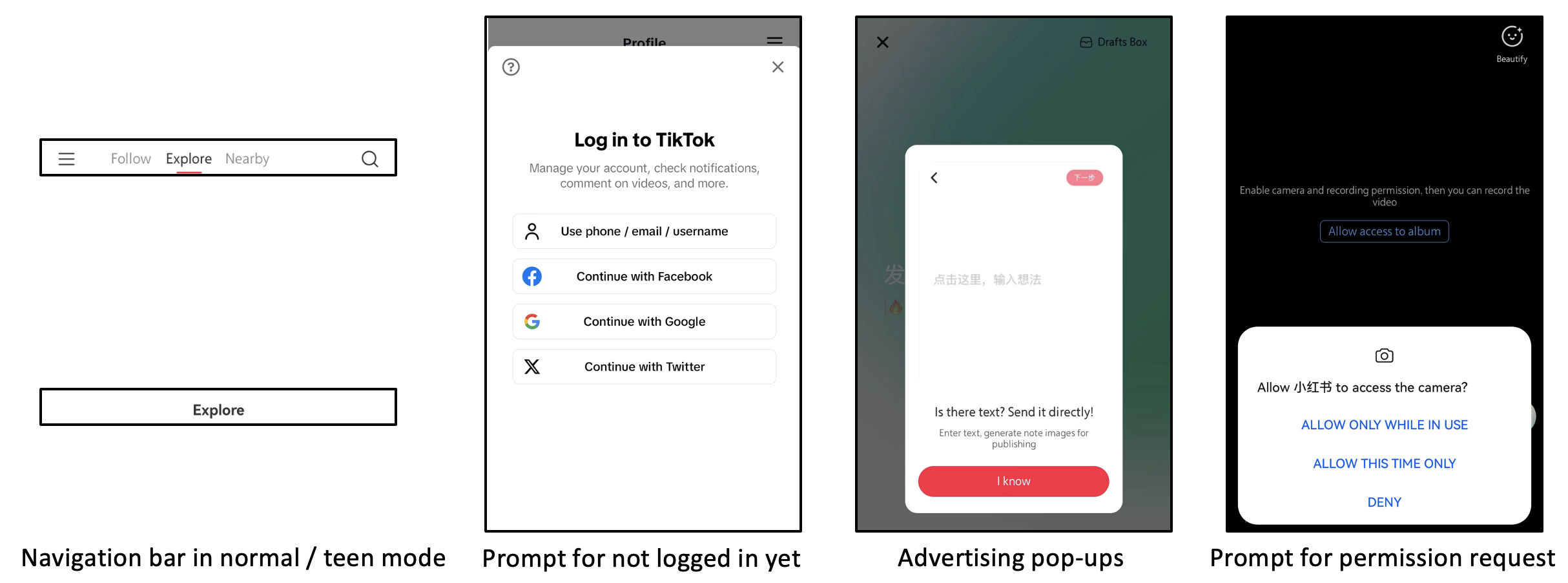}
  \caption{For the same textual prompt ``post a new moment'', however, the page state can affect the specific operations required.}
  \label{fig:pagestate}
   \vspace{3mm}
\end{figure}

\begin{table}[t]
	\centering
	\caption{\textcolor{revise}{Comparison to prior work. The comparison dimensions include the type of input content, whether the range of supported tasks is limited, whether the operation sequence can be updated in real-time based on the actual UI state, how to handle errors, and what types of knowledge are incorporated.}}
        \label{tab:relatedwork}
        \begin{threeparttable}
	\begin{tabular}{llcccl}
		\toprule  
		\multirow{2}{*}{Prior work}&\multirow{2}{*}{Input}&Unlimited &\multirow{2}{*}{Update}&Error&\multirow{2}{*}{Knowledge}\\
  &&Tasks&&Handling&\\
		\midrule  

            UIBert \cite{ijcai2021p235} & Single step &  \checkmark & $\times$ & ED\&R& Prepared dataset \\
            
            Spotlight \cite{li2023spotlight} & Single step & \checkmark & $\times$ & ED\&R & Prepared dataset \\

            \textcolor{minorrevise}{Ferret-UI} \cite{10.1007/978-3-031-73039-9_14, li2024ferretui2masteringuniversal} & Single step & \checkmark & $\times$ & ED\&R & Prepared dataset \\

            \multirow{2}{*}{MUG\cite{li-etal-2024-mug}} & \multirow{2}{*}{Single step}&\multirow{2}{*}{\checkmark}&\multirow{2}{*}{$\times$}&\multirow{2}{*}{UI}& Prepared dataset,\\
        &&&&&interaction history\\

   AdbGPT \cite{feng2023prompting} &Steps&\checkmark&$\times$&ED\&R& LLM\\

            Voicify \cite{10.1145/3581998}&Target Page+Steps&$\times$\tnote{1}& $\times$& ED\&R&  Prepared dataset\\
            
            UGIF \cite{venkatesh_ugif_2022}, \textcolor{minorrevise}{Macro} \cite{10.1145/3613904.3642074} &Function&$\times$\tnote{2}&$\times$&ED\&R& Prepared dataset\\

            AXNav \cite{Taeb_2024}&Function&\checkmark&\checkmark&ED\&R&LLM\\

            

            AutoDroid \cite{wen2024autodroid}&Function& \checkmark & \checkmark & ED\&R &Prepared UI graph\\
            
            \textcolor{minorrevise}{GPTVoiceTasker} \cite{Vu_2024} & Function & \checkmark & \checkmark & UI & Expanding UI Graph\\
            
            RTA \cite{zhang2023responsible}&Task-related texts&\checkmark&\checkmark&ED\&R&LLM\\

            CogAgent \cite{hong2023cogagent} & Task-related texts&\checkmark&\checkmark&ED\&R& Fine-tuned LLM\\

            \textcolor{minorrevise}{VisionTasker} \cite{10.1145/3654777.3676386} & Task-related texts & \checkmark & \checkmark & ED\&R & LLM, prepared dataset \\

            \textcolor{minorrevise}{Mobile-Agent} \cite{wang2024mobileagentautonomousmultimodalmobile} & Task-related texts & \checkmark & \checkmark & ED\&R & LLM \\ 
            \midrule
            \multirow{2}{*}{Prompt2Task} & \multirow{2}{*}{Task-related texts} & \multirow{2}{*}{\checkmark} & \multirow{2}{*}{\checkmark} & \multirow{2}{*}{UI}& Online tutorials,\\
            &&&&&accumulated knowledge\\
		\bottomrule  
	\end{tabular}
     \begin{tablenotes}
     \item ED\&R: Extend the dataset and retrain models to handle errors.
     \item UI: User intervention.
     \item{1} Only pages with deep links are supported.
     \item{2} Only functions in the dataset are supported.
    \end{tablenotes}
\end{threeparttable}
\end{table}

\textcolor{revise}{\textbf{2. Output:} The output environment for task automation is dynamic, which means that the operation sequence ($o_{1:n}$) depends not only on the textual instructions ($i_{1:n}$) but also on the operating environment. Factors like smartphone model \cite{10.1007/978-3-642-34478-7_69}, app version \cite{venkatesh_ugif_2022}, and page state \cite{10.1145/3610929} can all influence the operation sequence. For example, as shown in Figure \ref{fig:pagestate}, when a user wants to post a moment, varying page states may require different operations.} 




\textcolor{revise}{\textbf{3. Error Handling:} Successfully executing a multi-step automation task requires that each step be error-free, which is a strict condition. Besides the two challenges mentioned above, many other factors can lead to errors, such as missing accessibility labels on GUIs \cite{10.1145/3368089.3417940, 10.1145/3293882.3330551, wang_enabling_2023}, inadequate semantic recognition of GUI elements \cite{10.1145/2556288.2556979, PASSINI2008521, 10.1145/3242587.3242650, lee_pix2struct_2022}, and missing parameters during execution \cite{ni2020natural}. These issues can result in unpredictable outcomes and significantly compromise the reliability of the task automation system. While we acknowledge that data-driven models have the potential to improve accuracy with sufficient training data in the future, we are more eager to optimize the model's performance in a controlled and targeted manner for current usage scenarios. Usually, a simple and minor correction is sufficient to ensure the success of a task \cite{li-etal-2024-mug}.}

\textcolor{minorrevise}{\textbf{4. Knowledge:} Since the process requires the infusion of knowledge, ensuring reliable knowledge acquisition is critical. While LLMs are promising for planning and reasoning, they often lack the necessary domain-specific knowledge, leading to errors or hallucinations \cite{10.1145/3571730}, especially with uncommon or deeply hidden functions in applications \cite{Vu_2024}. Additionally, LLMs are often inefficient \cite{wang_enabling_2023}, making the automation process even slower than manual execution, which users find frustrating \cite{10.1145/2702123.2702588}. Recent works have started to recognize the importance of domain-specific knowledge, such as continuously constructing UI graphs to enhance task planning \cite{Vu_2024}. We believe a more systematic and comprehensive application of domain-specific knowledge is needed---one that ensures the knowledge required for task automation is not only accurate but also efficiently extracted and applied, improving both reliability and speed throughout the entire process.}


\textcolor{revise}{Considering the aforementioned challenges, prior work has proposed various solutions, each tackling some aspects of these issues to varying degrees. As shown in Table \ref{tab:relatedwork}, we identify significant room for improvement in their practical effectiveness when applied in real-world scenarios. To this end, we have designed and implemented Prompt2Task, which provides targeted solutions to the challenges outlined above.}

\section{Prompt2Task}

\begin{figure}[b]
  \centering
  \includegraphics[width=\linewidth]{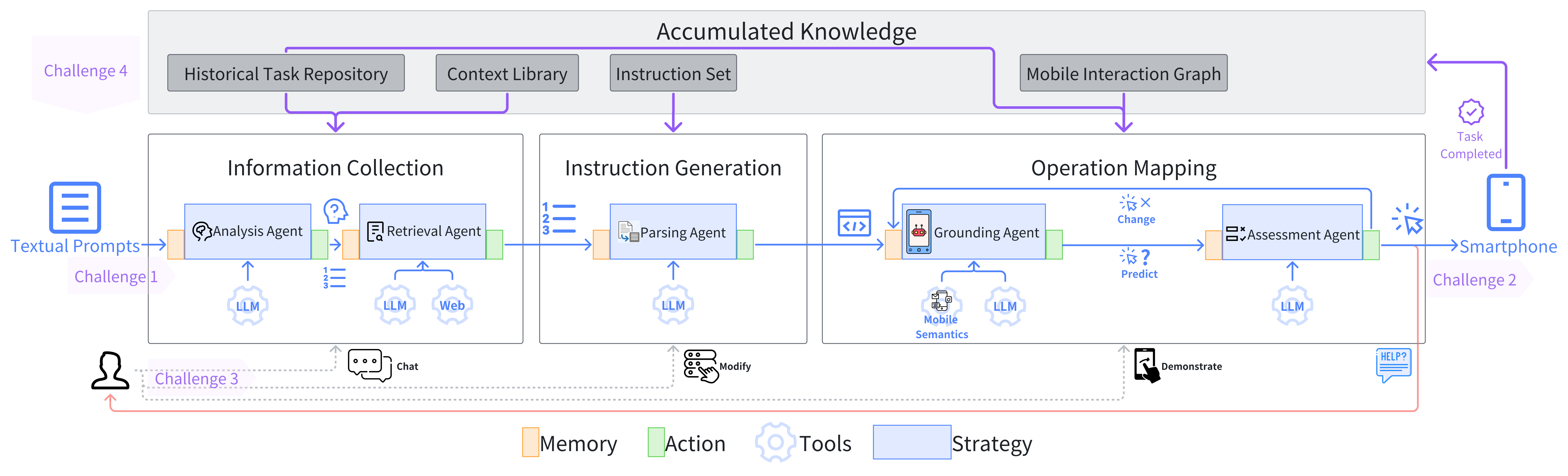}
  \caption{System design of Prompt2Task. The blue solid lines represent the main flow of the task automation process, the grey dashed lines represent optional user intervention, the purple lines represent knowledge accumulation after task completion and its optimization to each stage, and the red solid line denotes instances where the agent seeks user assistance. \textcolor{revise}{Each stage's workflow details are shown in Figure \ref{fig:informationcollection}, \ref{fig:instructiongeneration}, \ref{fig:operationmapping}.}}
  \label{fig:systemdesign}
\end{figure}

\textcolor{revise}{To address the above challenges and align with the components of the Prompt2Task model, we propose Prompt2Task, as shown in Figure \ref{fig:systemdesign}. We have divided the process into three stages:}

\textcolor{revise}{\textbf{Information Collection}: To tackle the challenge posed by inputs, Prompt2Task extracts useful information and underlying intentions from the textual prompt $P$. It supplements missing information in the user's expression through predictive analysis, asking the user or retrieving external knowledge. This stage prepares all relevant natural language information: the function description $F$ and step description $s_{1:n}$.}  


\textcolor{revise}{\textbf{Instruction Generation}: Employing the collected information, this stage strategizes the task's progression. Given the instruction definition, Prompt2Task converts the step description into a list of standardized instructions \(i_{1:n}\).} 


\textcolor{revise}{\textbf{Operation Mapping}: To address challenges from the dynamic output environment, such as rapidly updated pages, Prompt2Task should make adaptive decisions beyond pre-defined instructions. Specifically, it includes a deep understanding of the real-time environment, flexible grounding, and assessment to enhance the reliability of operations. This stage finalizes the generation of the operation sequence $o_{1:n}$.}


\textcolor{revise}{To optimize the system in a controlled and targeted manner, Prompt2Task is designed to seek user assistance when errors or uncertainties arise, ensuring the task remains on track. The trade-off here is balancing user oversight to enhance system effectiveness while minimizing user burden. Prompt2Task achieves this by making the results of each stage transparent and requesting user help only when necessary. Even if users are not willing to intervene, Prompt2Task can autonomously complete the entire process.}



\textcolor{revise}{In terms of knowledge, we incorporate online tutorials to expand the task range as open-ended as possible and enhance the reliability of planning. The accumulated knowledge from past tasks, including context and execution information, is processed and stored to improve Prompt2Task's future reasoning capabilities and efficiency.}

\textcolor{revise}{Throughout the above three stages, different types of intelligence and knowledge are applied to achieve the corresponding goals. Therefore, we utilize a multi-agent system framework. Each agent, based on their assigned sub-tasks, is responsible for specific knowledge retrieval, tool utilization, decision-making, and action execution. These agents collaborate to complete the overall task. Figure \ref{fig:systemdesign} illustrates how Prompt2Task is partitioned into specialized agents and a unified knowledge base. We will delve into each in the following subsections.}

\subsection{Multi-Agent Design}





Drawing inspiration from the prior agent system frameworks \cite{panait2005cooperative,8352646,talebirad2023multiagent,hong2023metagpt}, each agent within Prompt2Task is composed of several key components: memory, tools, strategy, and actions, which support the agent's fundamental functions. Additionally, each agent is equipped with a user interface that allows users to observe, modify, or augment agent decisions, as shown in Figure \ref{fig:systemdesign}. The interface is not just for monitoring but aims to foster human-agent collaboration.  

We have designed five distinct agents specifically tailored to the three stages of the process.

\begin{figure}[htbp]
  \centering
  \includegraphics[width=\linewidth]{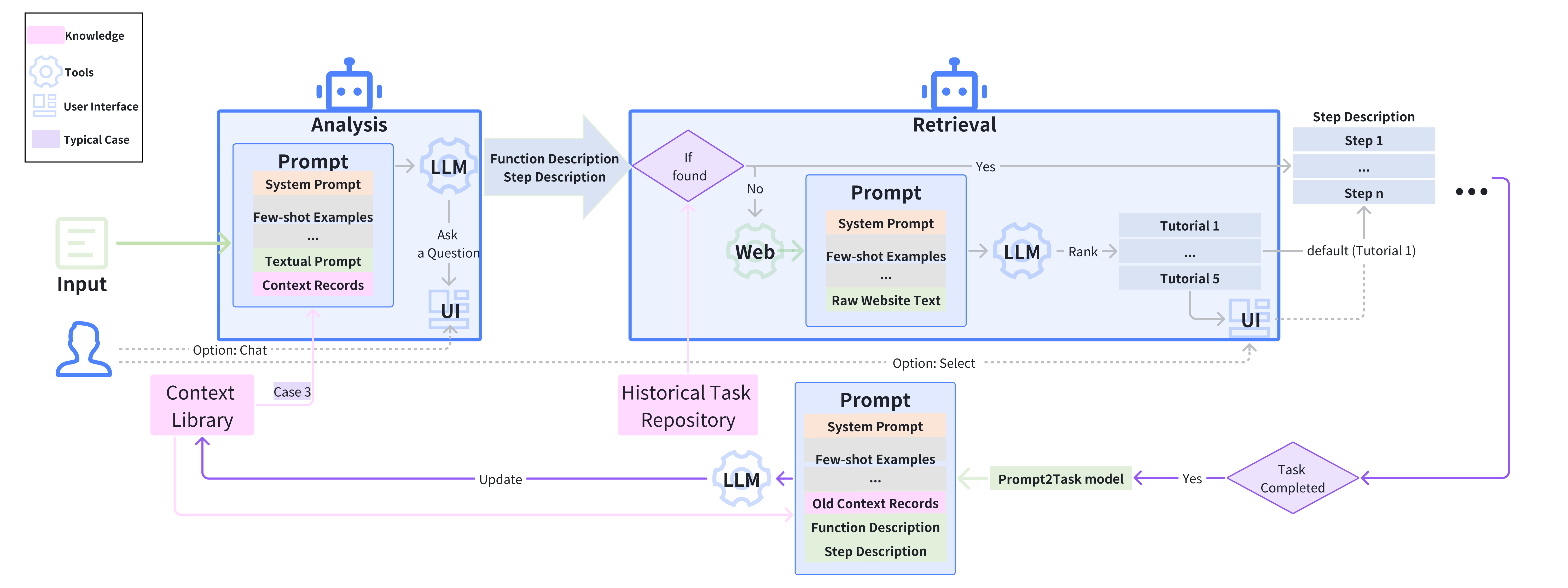}
  \caption{\textcolor{revise}{Information collection workflow. Task-related functions and step descriptions are collected by the \emph{Analysis Agent} and \emph{Retrieval Agent}. The colors of the components within the prompt indicate their sources, which are the same color elements pointing to the prompt with the same color arrows. Gray arrows represent the normal task flow, while purple arrows indicate the accumulation of knowledge.}}
  \label{fig:informationcollection}
\end{figure}

\begin{itemize}
\item \textbf{Information Collection} (Figure \ref{fig:informationcollection}):
\begin{itemize}
\item \textit{Analysis Agent}: Extracts and analyzes information implicit in the textual prompt to construct a complete \textbf{function description $F$}.
\item \textit{Retrieval Agent}: Acquires external knowledge relevant to the function description and builds an enriched \textbf{step description $s_{1:n}$}.
\end{itemize}
\item \textbf{Instruction Generation} (Figure \ref{fig:instructiongeneration}):
\begin{itemize}
\item \textit{Parsing Agent}: Transforms the collected information into a list of \textbf{instructions $i_{1:n}$}.
\end{itemize}
\item \textbf{Operation Mapping} (Figure \ref{fig:operationmapping}):
\begin{itemize}
\item \textit{Grounding Agent}: Predicts \textbf{operations $o_{1:n}$} on the smartphone based on the generated instructions.
\item \textit{Assessment Agent}: Reviews the predicted operations, evaluates if user intervention is required, and eventually executes operations on smartphones.
\end{itemize}
\end{itemize}

Subsequently, we will introduce how each agent works.



\subsubsection{Analysis Agent}
\label{sec:analysisagent}


\textcolor{revise}{The agent aims to analyze and confirm task-related information from the textual prompts.} 

The prompts may include descriptions of the function or specific steps in the model, as outlined in Section \ref{sec:rpadefinition}. \textcolor{revise}{The function description must be confirmed and refined from the input.} The original function descriptions are often inforaml and may lack essential context, such as specific application names or phone models. Figure \ref{fig:dialog} demonstrates how the agent interacts with users to refine these descriptions. Utilizing the LLM, the agent successively adopts three key strategies:

\begin{figure}[b]
  \centering
  \includegraphics[width=\linewidth]{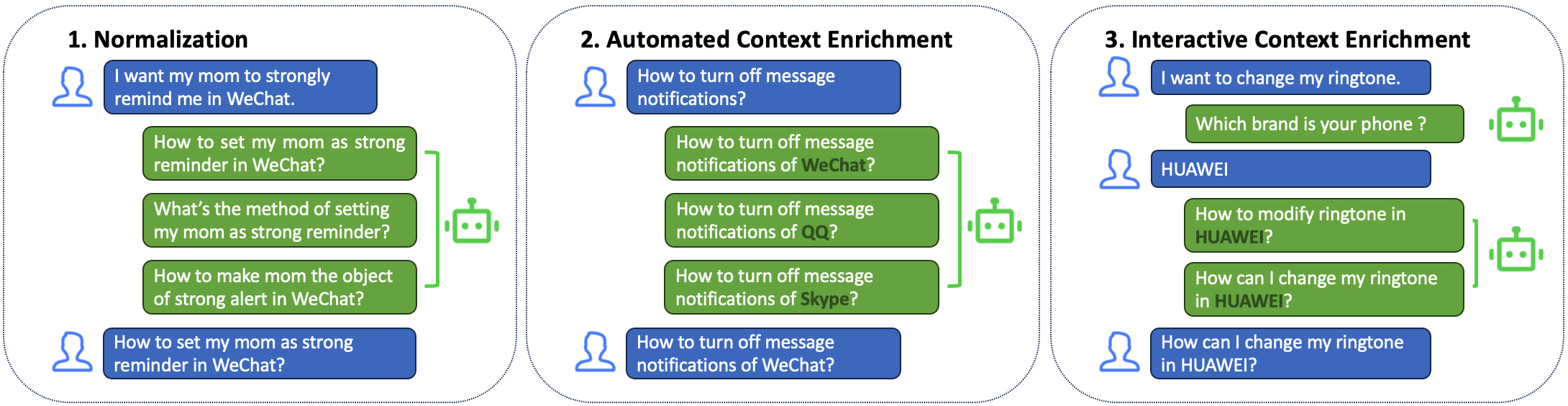}
  \caption{Three scenarios where the agent interacts with the user to refine the function description.}
  \label{fig:dialog}
\end{figure}

\begin{itemize}
\item \textbf{Normalization}: \textcolor{revise}{Users often describe the desired outcome without accurately naming the specific function. To address this, the agent rephrases the original input to reduce colloquial and inaccurate expressions.}

\item \textbf{Automated Context Enrichment}: \textcolor{revise}{The agent utilizes the LLM to enhance the user's original input by incorporating the user's record from the context library (detailed in Section \ref{sec:contextlibrary}). The record is directly added in the prompt to the LLM, enabling the generation of several potential candidates for the user to choose from.}

\item \textbf{Interactive Context Enrichment}: \textcolor{revise}{If the agent determines that relevant context information is still missing after the automated enrichment, it queries the user to obtain the missing details. The agent continues to chat with the user until it determines that the function description is sufficiently clear.}

\end{itemize}

\textcolor{revise}{After refining and enriching the function description, the user confirms it via chatting (Figure \ref{fig:prototype1}).} The agent then bundles the function description, along with step descriptions (if available), and forwards them to the \emph{Retrieval Agent} for subsequent processing.





\subsubsection{Retrieval Agent}



The \emph{Analysis Agent} can ensure a clear function description, but it may fall short when it comes to providing detailed step descriptions, which the prompt might not always cover. Addressing this void, the \emph{Retrieval Agent} is responsible for preparing a complete step description. For efficiency, the agent prioritizes the following strategies:

\begin{itemize}
    \item If the user confirms that the step description in the prompt is complete, the system advances to the next stage without additional information collection.
    \item Otherwise, the agent searches the historical task repository (detailed in Section \ref{sec:historicalrpa}) for matching the most relevant historical task model, using function description as the keyword. If a match is found, the agent will use the matched task model, which has been previously used within our system and is deemed reliable.

\end{itemize}

If the user is dissatisfied or no match is found, the agent searches for relevant tutorials on the Internet. The agent uses search engines, with the function description as a query. The tutorial texts from the fetched web pages are refined using SimpRead\footnote{\href{https://simpread.pro/}{https://simpread.pro/}}'s rule library and summarized by the LLM. \textcolor{revise}{As illustrated in Equation \ref{eq:rankscore}, the score for each tutorial is calculated based on criteria that include their original search ranking, the data source, and textual quality assessed by the LLM \cite{chen2023exploring}. Notably, the alignment with the original step description analyzed from the user's prompt is also a factor to consider, which is also assessed by the LLM. The constant parameters are empirically determined. Tutorials are ranked according to their score from highest to lowest.}

\textcolor{revise}{
\begin{equation}
\label{eq:rankscore}
\begin{aligned}
Score(Tutorial_i) &= c_0 \times originalRank_i 
\\&+ c_1 \times weight(source_i) 
\\&+ c_2 \times textQuality_i
\\&+ c_3 \times matched(i,oriStepDescription)
\end{aligned}
\end{equation}}

\textcolor{revise}{As shown in Figure \ref{fig:prototype2}, Prompt2Task presents the top 5 highest-ranking tutorials to users, automatically selecting the highest-ranked tutorial in cases where users do not make a choice.} By this stage, both the function and step description for the task automation are prepared in place.




\begin{figure}[htbp]
  \centering
  \includegraphics[width=\linewidth]{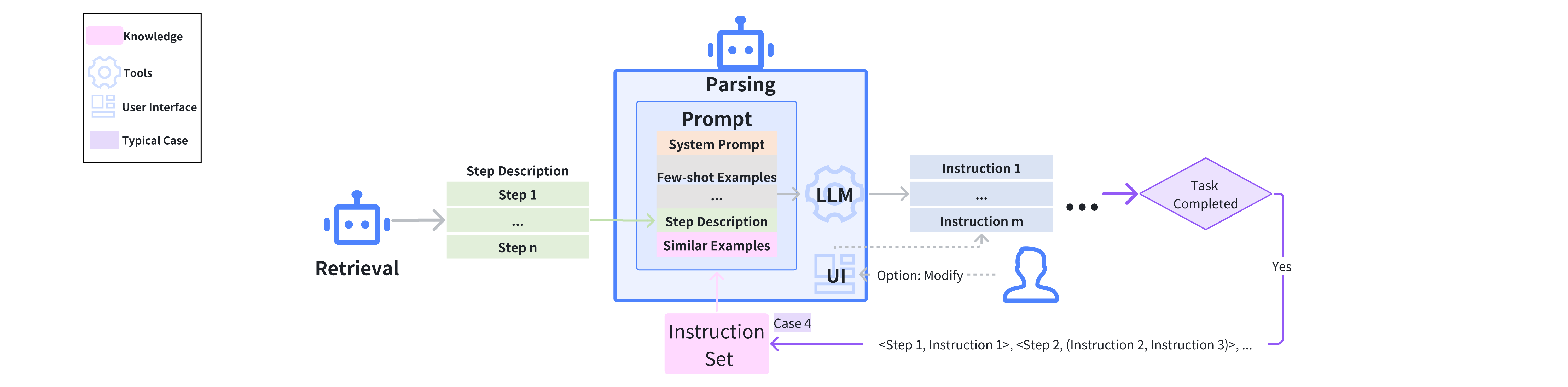}
  \caption{\textcolor{revise}{Instruction generation workflow. Task-related instructions are generated by the \emph{Parsing Agent}. The colors of the components within the prompt indicate their sources, which are the same color elements pointing to the prompt with the same color arrows. Gray arrows represent the normal task flow, while purple arrows indicate the accumulation of knowledge.}}
  \label{fig:instructiongeneration}
\end{figure}

\subsubsection{Parsing Agent}
The agent's responsibility is to translate step descriptions into a sequence of formalized instructions tailored for smartphone operations. Building on previous work \cite{li_mapping_2020}, the definition of instruction has been enriched to include a more detailed object description. As shown in Table \ref{tab:operation}, an instruction is represented as a tuple comprising the operation type, associated parameters, and object descriptions, usually referencing a widget on the page. The agent utilizes LLM for parsing, because of its capacity to function effectively with minimal initial data and its adeptness in handling flexible text types (e.g., natural language, several keywords). As the number of cases increases, the instruction definition can be extended and the agent can be progressively refined (further discussed in Section \ref{sec:instructionset}). \textcolor{revise}{As shown in Figure \ref{fig:prototype3}, users can validate the parsed results and, where necessary, can perform manual corrections.}

\begin{table}[htbp]
	\centering
	\caption{Formalized definitions of instructions.}
        \label{tab:operation}
	\begin{tabular}{lll}
		\toprule  
		Operation&Parameter&Object\\ 
		\midrule  
            open& application name&\multirow{6}{4cm}{description of UI pages\\ or widgets\\ (text, position, color...) }\\
		click& times (default=1)&\\
            long click&time&\\
            switch&value (true, false)&\\
            edit&text&\\
            scroll&direction (up, down, ...)&\\
            back&page (previous, homepage)&\\
		\bottomrule  
	\end{tabular}
\end{table}




\begin{figure}[htbp]
  \centering
  \includegraphics[width=\linewidth]{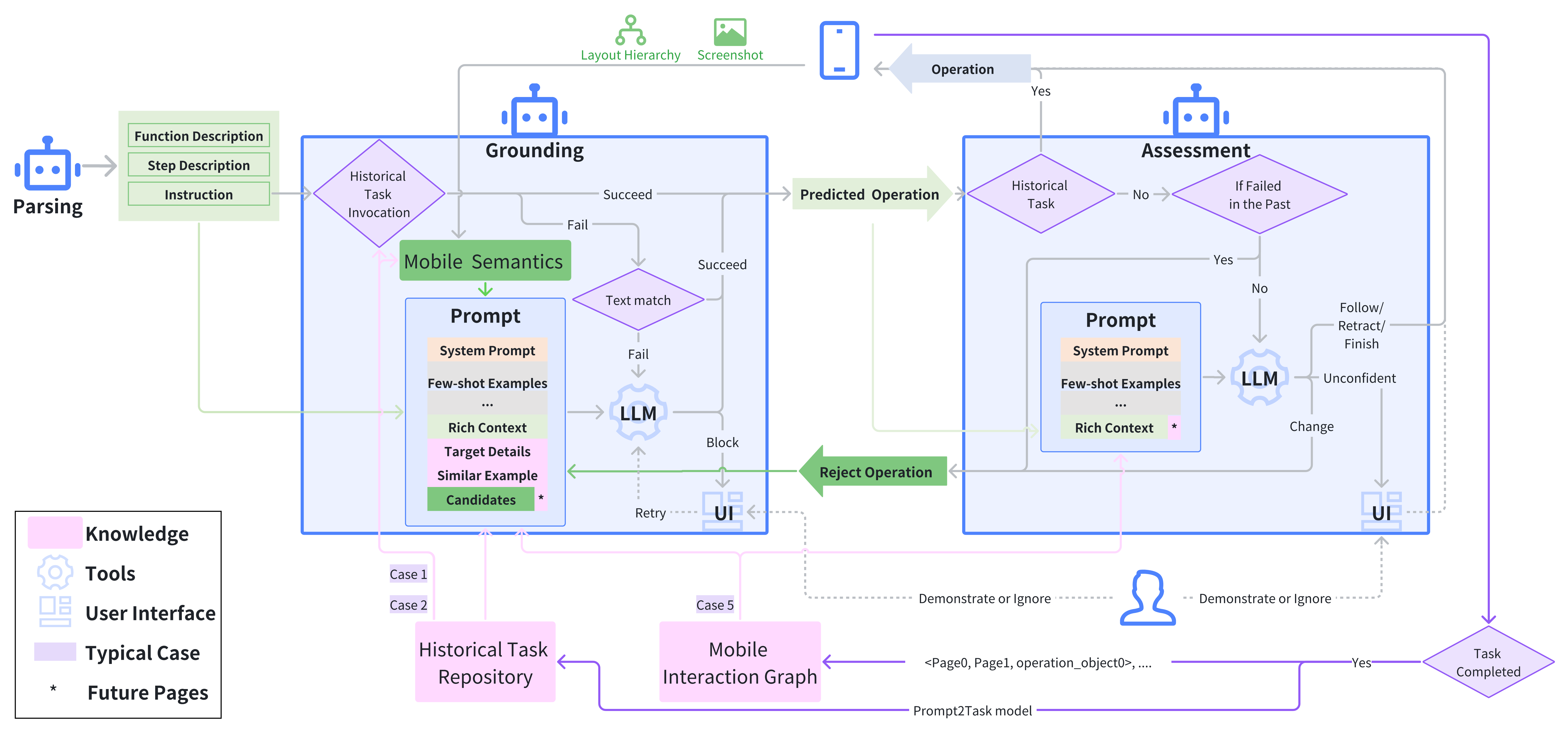}
  \caption{\textcolor{revise}{Operation mapping workflow. The \emph{Grounding Agent} predicts the operation and the \emph{Assessment Agent} performs the final check. The colors of the components within the prompt indicate their sources, which are the same color elements pointing to the prompt with the same color arrows. Gray arrows represent the normal task flow, while purple arrows indicate the accumulation of knowledge.}}
  \label{fig:operationmapping}
\end{figure}

\subsubsection{Grounding Agent}
\label{sec:grounding}



As shown in Figure \ref{fig:operationmapping}, the final stage aims to predict specific operations on the smartphone given the instructions. Specifically, according to the object description in the instruction, it identifies the UI widget on the smartphone and thus executes the relevant operation.


\textcolor{toconfirm}{Initially, the agent filters out all widgets that lack both interactive attributes and text descriptions. Interactive attributes include being clickable, long clickable, editable, focusable, checkable, and scrollable. Text descriptions include both explicit text and implicit content descriptions. They are all available in the layout hierarchy.} Then the agent relies on the mobile semantics module to bridge the gap between text-based instructions and GUI described by layout hierarchy and screenshots. Considering real-time efficiency, the module employs the \textit{screen2words} model \cite{wang_screen2words_2021} to generate full-page textual descriptions and utilizes a combination of OCR and \textit{pix2struct} \cite{pmlr-v202-lee23g} to describe widgets which lack text descriptions.

Subsequently, to predict the operation efficiently and effectively, the agent adopts a multi-layer strategy as follows:


\textbf{Historical task invocation}: If the given instructions are recurring, the agent consults the historical task repository (in Section \ref{sec:historicalrpa}) to identify widgets used in the past. It then compares these widgets with the ones on the current page, using heuristic rules that compare their rich GUI attributes. A widget is considered a match only if it exceeds the preset similarity thresholds. 

\begin{figure}[htbp]
	\centering
	\subfigure[]{
		\begin{minipage}[b]{0.48\textwidth}
		\centering
			\includegraphics[height=2.4cm, keepaspectratio]{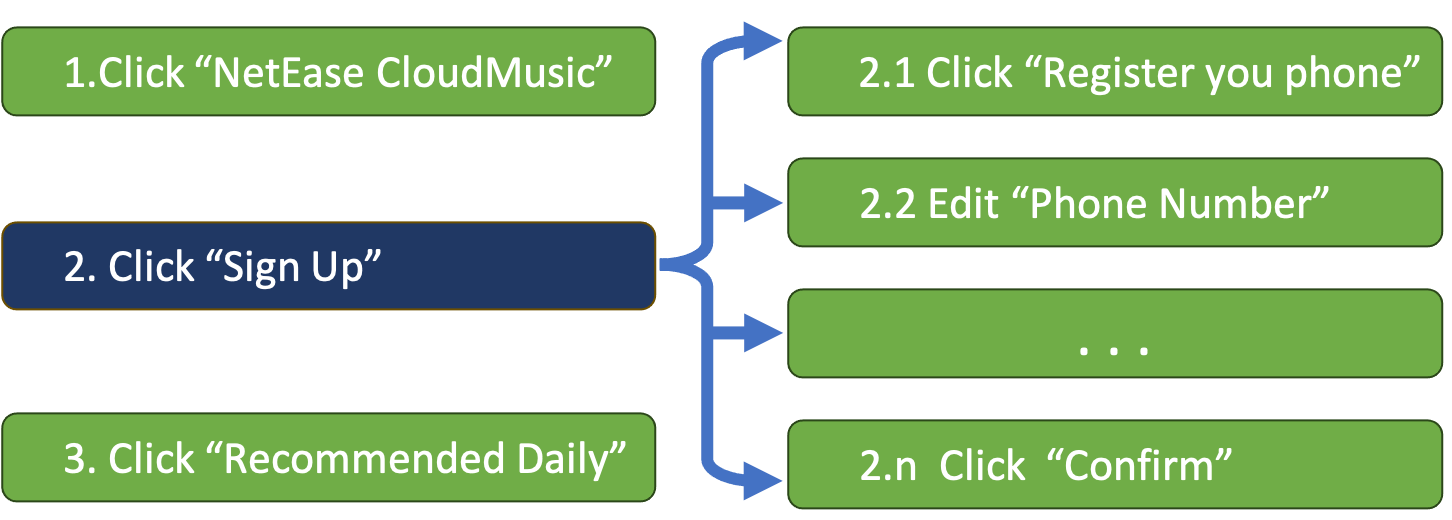}
		\end{minipage}
		\label{fig:nest}
	}
    \subfigure[]{
            \begin{minipage}[b]{0.48\textwidth}
		\centering
		\includegraphics[height=2.4cm, keepaspectratio]{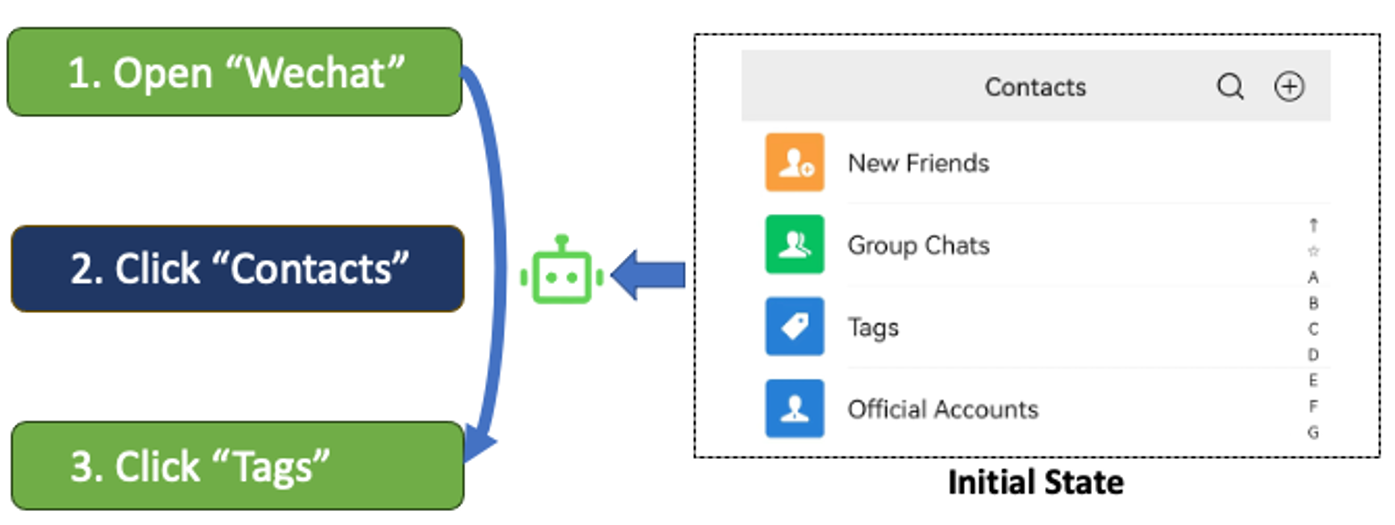}
              \end{minipage}
              \label{fig:skip}
    }
        \caption{(a) The compound instruction \emph{Click ``Sign up''} requires additional instructions. The agent can add the necessary steps. (b) The   state doesn't match the first instruction. The agent can skip unnecessary instructions.}
\end{figure}

\textbf{New instruction-based grounding}: \textcolor{revise}{For new instructions or when the historical task invocation fails}, the agent employs the following strategies:
    \begin{enumerate}
        \item \textbf{Strict text matching}: The agent uses strict text matching to find a widget closely resembling the object description. A preset similarity threshold must be met to confirm a match. 

        \textcolor{toconfirm}{If this strategy fails, the agent then employs the LLM to decide which of the remaining five strategies to use: redirect, add, skip, expand, and block.}
        \textcolor{toconfirm}{\item \textbf{Redirect}: The agent leverages LLM to predict the widget most suited to execute the given instruction. It focuses on semantic similarity with the operation and object description in the instruction.} 
        
        \textcolor{toconfirm}{However, instructions and operations do not always correspond one to one. To make the mapping more flexible, the agent can choose from the following exploratory strategies:}
        \textcolor{toconfirm}{\item \textbf{Add}: 
        The agent incorporates additional steps when necessary, often due to outdated or erroneous instructions, or when an instruction step requires expansion (Figure \ref{fig:nest}). The agent leverages as much contextual information as possible to infer the most probable operation rather than solely relying on the initial instructions. 
When the agent predicts an incorrect operation, the \emph{Assessment Agent} can help it retract and potentially find the correct operation. This strategy is crucial when there are unexpected changes on the page (Figure \ref{fig:pagestate}).}
\textcolor{toconfirm}{\item \textbf{Skip}: The agent skips unnecessary or irrelevant steps to better align the current page state with the instructions. This strategy helps in addressing inconsistent starting points, allowing the process to continue smoothly even if some steps are not applicable (Figure \ref{fig:skip}).}
\textcolor{toconfirm}{\item \textbf{Expand}: When all instructions have been finished, or the remaining instructions are totally useless, the agent keeps trying to explore potential operations, until the \emph{Assessment Agent} indicates task completion.}
\textcolor{toconfirm}{\item \textbf{Block}: The agent is currently unable to proceed as it cannot identify a suitable interactive widget, often due to the lack of information on the current page. In such cases, the agent requests user intervention by describing the current situation and the expected user actions. The user is asked to manually demonstrate the correct operation on the mobile GUI, as shown in Figure \ref{fig:prototype4}. If the user chooses to ignore, the agent will re-invoke the LLM to employ one of the above strategies.}
    \end{enumerate}

\textcolor{toconfirm}{As the agent persistently accrues knowledge and insights, it improves mobile semantics extraction and LLM reasoning (further discussed in Section \ref{sec:interactiongraph}). For the implementation details of each strategy, see Appendix \ref{apd:groundingagent}.}








\subsubsection{Assessment Agent}



The agent is introduced to proactively address potential pitfalls and simplify decision-making. It ensures the \emph{Grounding Agent} remains focused on its primary tasks without being overloaded with error detection.

\textcolor{toconfirm}{As shown in Figure \ref{fig:operationmapping}, for efficiency, historical task invocations are thought to be sufficiently reliable and thus do not require assessment, allowing for direct execution. For new instruction-based grounding, the agent first assesses whether the current operation is a previously failed attempt, preventing endless loops during exploration.} Subsequently, the agent evaluates whether the operations executed up to the present are reasonable. The agent leverages the LLM for reasoning. Based on contextual information including the current page description, operation history, and the current instruction, the agent adopts one of the following strategies:

\begin{itemize}
    \item \textcolor{toconfirm}{\textbf{Follow}: The agent determines that the operation is reasonable and proceeds with execution on the smartphone.}
    \item \textcolor{toconfirm}{\textbf{Change}: The agent determines that the operation does not meet the current task requirements and instructs the \emph{Grounding Agent} to remove the erroneous option and make a new prediction.}
    \item \textcolor{toconfirm}{\textbf{Retract}: If the \emph{Grounding Agent} has previously attempted erroneous expansions leading to irrelevant pages, the agent performs a global back operation to return to the previous step.}
    \item \textcolor{toconfirm}{\textbf{Finish}: Usually, tasks are not completed simply by executing all instructions, due to outdated tutorials or errors in preceding stages. Therefore, the \emph{Grounding Agent} will automatically choose the ``expand'' strategy to identify further necessary operations after all instructions have been executed. Based on the contextual information, the \emph{Assessment Agent} determines whether the task is actually completed. The execution process only terminates when the \emph{Assessment Agent} confirms task completion.}
\end{itemize}



\begin{figure}[t]
  \centering
  \includegraphics[width=.5\linewidth]{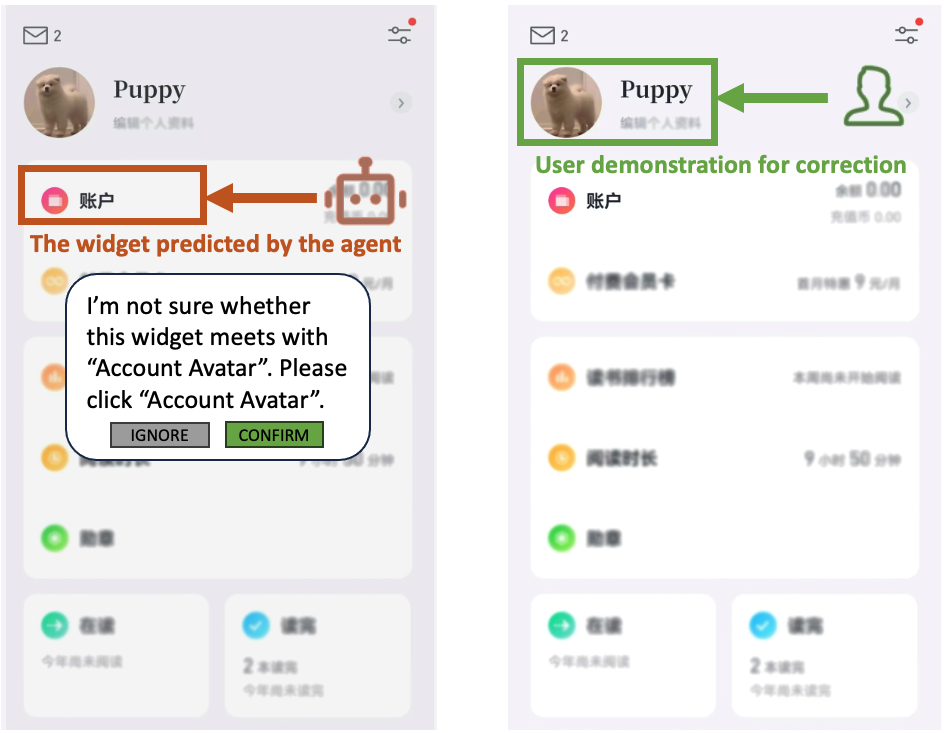}
  \caption{A case of decision uncertainty prompting user intervention. The target object is described as ``Account Avatar''. The agent, which relies heavily on textual descriptions for predictions, inaccurately predicts ``account'' because the avatar widget lacks a precise textual description. Due to low confidence in its prediction, the agent seeks user intervention.}
  \label{fig:demonstrate}
\end{figure}

\textcolor{toconfirm}{However, sometimes the agent is not sufficiently confident in adopting any of the above strategies. In such cases, we consider it appropriate to seek user assistance. Therefore, the agent employs the LLM to assess if it is confident in its decision. If the agent is unconfident, it describes its current decision and prompts the user to demonstrate the correct operation. Figure \ref{fig:demonstrate} illustrates a scenario where the agent prompts user intervention due to low confidence for decision. Similarly, Figure \ref{fig:title} showcases a situation where the instruction ``Click a ringtone you like'' introduces an ambiguity that necessitates user assistance. More details are in Appendix \ref{apd:assessmentagent}.}

Notably, even if the agent prompts to seek user assistance, the user has the option to ignore the prompt, in which case the agent will directly execute according to its own decision.



\subsection{Accumulated Knowledge}

As shown in Figure \ref{fig:informationcollection}, \ref{fig:instructiongeneration}, \ref{fig:operationmapping}, the agents draw from an accumulated knowledge base, which is continually improved through user usage and intervention. As a feedback mechanism, the knowledge base facilitates the learning of the agents, thereby enhancing the entire system's efficiency and task success rates.

\textcolor{revise}{According to the above agents' knowledge requirements, Prompt2Task requires four types of knowledge:}

\begin{itemize}
    \item \emph{Historical Task Repository} stores successfully executed automation tasks, which typically exhibit a higher success rate in future runs. Additionally, it retains adjustments tailored to specific operating environments or updates to previously outdated tutorials, eliminating the need for manual reconfigurations upon reuse.

    \item \emph{Context Library} stores and continuously updates user-specific mobile parameters. It helps the system supplement necessary details when faced with incomplete user expressions.

    \item \emph{Instruction Set} focuses on improving the system's ability to extract instructive information from text. Similar examples become valuable learning materials.

    \item \emph{Mobile Interaction Graph} strengthens the system's understanding of app functionalities and navigation logic. It serves as an auxiliary routing tool, helping agents to better predict the operations.

\end{itemize}

We will introduce them in details and explain how they bring improvements to the agents.



\subsubsection{Historical Task Repository} 
\label{sec:historicalrpa}

\begin{sloppypar}
Upon successful execution, Prompt2Task archives the Prompt2Task model's data, according to the definition in Section \ref{sec:rpadefinition}. Since the \emph{Grounding Agent} may not always align operations with the original instructions, the repository uses the sequence of operations as the ground truth. These operations are then reverse-engineered into corresponding instructions and step descriptions via simple mapping rules, as detailed in Appendix \ref{apd:historicalrpa}.
\end{sloppypar}


The historical task repository provides several optimizations to the agents:

\begin{itemize}
    \item \textcolor{revise}{The \emph{Retrieval Agent} will prioritize historical task models, enhancing data reliability and avoiding the time on retrieving online tutorials whenever possible. The agent identifies the most similar historical task models based on the current function description. Only when the similarity exceeds the preset empirical threshold is the model considered a match and directly used for subsequent processing.}
    \item \textcolor{toconfirm}{The \emph{Grounding Agent} gains enhanced predictive capabilities. For instructions stored in the repository, it can now identify corresponding UI widgets through a richer feature set that incorporates layout hierarchies and screenshots, rather than relying solely on text descriptions from instructions (as exemplified in Case 1 in Appendix \ref{apd:cases}). As long as the features of the UI widgets do not change significantly, the heuristic rules employed by the agent can accurately and quickly identify the widgets and trigger the appropriate operations. Additionally, the stored mapping between instructions and operations serves as a learning resource for the agent. If the historical task invocation fails, the agent finds the historical instruction most similar to the current one and then incorporates the corresponding mapping into the prompt, thereby aiding the LLM in decision-making (for details, refer to the prompt engineering of the \emph{Grounding Agent} in Appendix \ref{apd:groundingagent}).}
    

    \item The mobile semantics module in the \emph{Grounding Agent} benefits from a richer set of textual descriptions for widgets. \textcolor{toconfirm}{For the widgets' screenshots stored in the repository, the module computes their embeddings \cite{wang_screen2words_2021}. When new images with similar embeddings to the stored ones appear, the module uses the stored textual descriptions for them, which are more accurate. It is particularly effective for recognizing icons; as long as the icons' images maintain a similarity that does not exceed an empirical threshold, the agent can accurately label them when they are encountered again (as exemplified in Case 2 in Appendix \ref{apd:cases}).}
\end{itemize}

\subsubsection{Context Library} 
\label{sec:contextlibrary}

Derived from a formative study involving 24 college students and 20,265 online Q\&A questions (detailed in Appendix \ref{apd:contextlibrary}), the library classifies these parameters into three categories based on acquisition methods:

\begin{enumerate}
\item \textbf{Automated System Capture}: Parameters that are available from the mobile GUI, like battery status or current application in focus, are automatically collected by the system. Sensitive data is not stored. 
\item \textbf{User Input}: Device-specific information authorized or provided by the user is also collected. It could include permissions granted (e.g., access to camera or location), as well as other profile details (e.g., preferred language) which users can actively input through forms presented by the system.
\item \textbf{User-Driven Updates}: The library continuously updates based on user-system interactions, \textcolor{revise}{primarily involving parameters mentioned by users in historical tasks, such as connected devices or Wi-Fi names.}
\end{enumerate}



In the library, a unique set of parameter records is maintained for each user (refer to the example in Appendix \ref{apd:contextlibrary}). \textcolor{revise}{To support user-driven updates, after each completed automation task, the original records and the current task model are provided to the LLM to generate the latest records.}

\textcolor{revise}{The \emph{Analysis Agent} accesses the library to find the user's historical records and incorporates these records as part of the prompt (for details, refer to the prompt engineering of the \emph{Analysis Agent} in Appendix \ref{apd:analysisagent}).} With the aid of the library, missing parameters for recurring textual prompts are automatically supplemented, which users have the option to confirm. \textcolor{revise}{For instance, the earphones mentioned by the user in a certain use (refer to Case 3 in Appendix \ref{apd:cases}) will be collected into the library and automatically suggested to the user in future tasks. Additionally, if the user's context changes and the parameters saved in the context library become outdated, users can correct them while chatting with the \emph{Analysis Agent}, and the parameters in the context library will be updated after the completion of the new task.}



\subsubsection{Instruction Set} 
\label{sec:instructionset}



The \emph{Parsing Agent} is tasked with converting natural language step descriptions into formalized instructions. While conceptually straightforward, practical application introduces issues, including the emergence of new operation, parameter, or object types and synonyms or deviation from the format in Table \ref{tab:operation}. To address these issues, the instruction set is regularly updated to include misinterpreted cases and their corrections. These cases can be resolved by user manual corrections, but most are resolved by instructions reverse-engineered from successful automation tasks, as described in Section \ref{sec:historicalrpa}. \textcolor{toconfirm}{Similar to the historical task repository, the instruction set stores the mapping between the description of a single step and the instruction it is parsed into. For each step that needs parsing, the agent identifies the most similar description from the set, and incorporates the historical mappings into the prompt for the LLM, thereby enhancing the LLM's parsing capabilities.}


\textcolor{revise}{Another important issue is related to compound instructions (as exemplified in Figure \ref{fig:nest} and Case 4 in Appendix \ref{apd:cases}). For such cases, the user marks the compound instruction and its decomposed result, which are then recorded in the instruction set. Hence, if a compound instruction reoccurs, the \emph{Parsing Agent} will automatically decompose it.}

\subsubsection{Mobile Interaction Graph} 
\label{sec:interactiongraph}

In the graph, each node represents a visited mobile page, and edges link interactive widgets to the pages they lead to \cite{deka_rico_2017}. As the system navigates more pages, the graph evolves dynamically. \textcolor{toconfirm}{Specifically, for a completed automation task, the operation sequence naturally corresponds to a series of page transitions. To represent these transitions in the graph, Prompt2Task uses a heuristic rule-based similarity function \cite{10.1145/3610929} that considers the texts and their locations on pages, to calculate their similarity and categorize the pages. The similarity thresholds have been established using Receiver Operating Characteristic (ROC) curve analysis. Similar pages share the same node in the graph, and edges between these nodes represent the operation objects that trigger transitions between pages.}

The graph enables virtual validation of each operation before actually executing on a smartphone. If the edge already exists in the graph, it anticipates the future pages that could be visited after the operation, thus enriching the contextual understanding for the \emph{Grounding Agent} and the \emph{Assessment Agent} (see Appendix \ref{apd:groundingagent} and \ref{apd:assessmentagent} for prompt details). \textcolor{revise}{As long as the page transition relationships remain unchanged, the graph can provide the agents with accurate information about future pages (as exemplified in Case 5 in Appendix \ref{apd:cases}).}






\subsection{Implementation}
\label{sec:implementation}

\begin{figure}[t]
    \centering

\subfigure[]{
		\begin{minipage}[t]{0.18\textwidth}
		\centering
			\includegraphics[height=4.6cm, keepaspectratio]{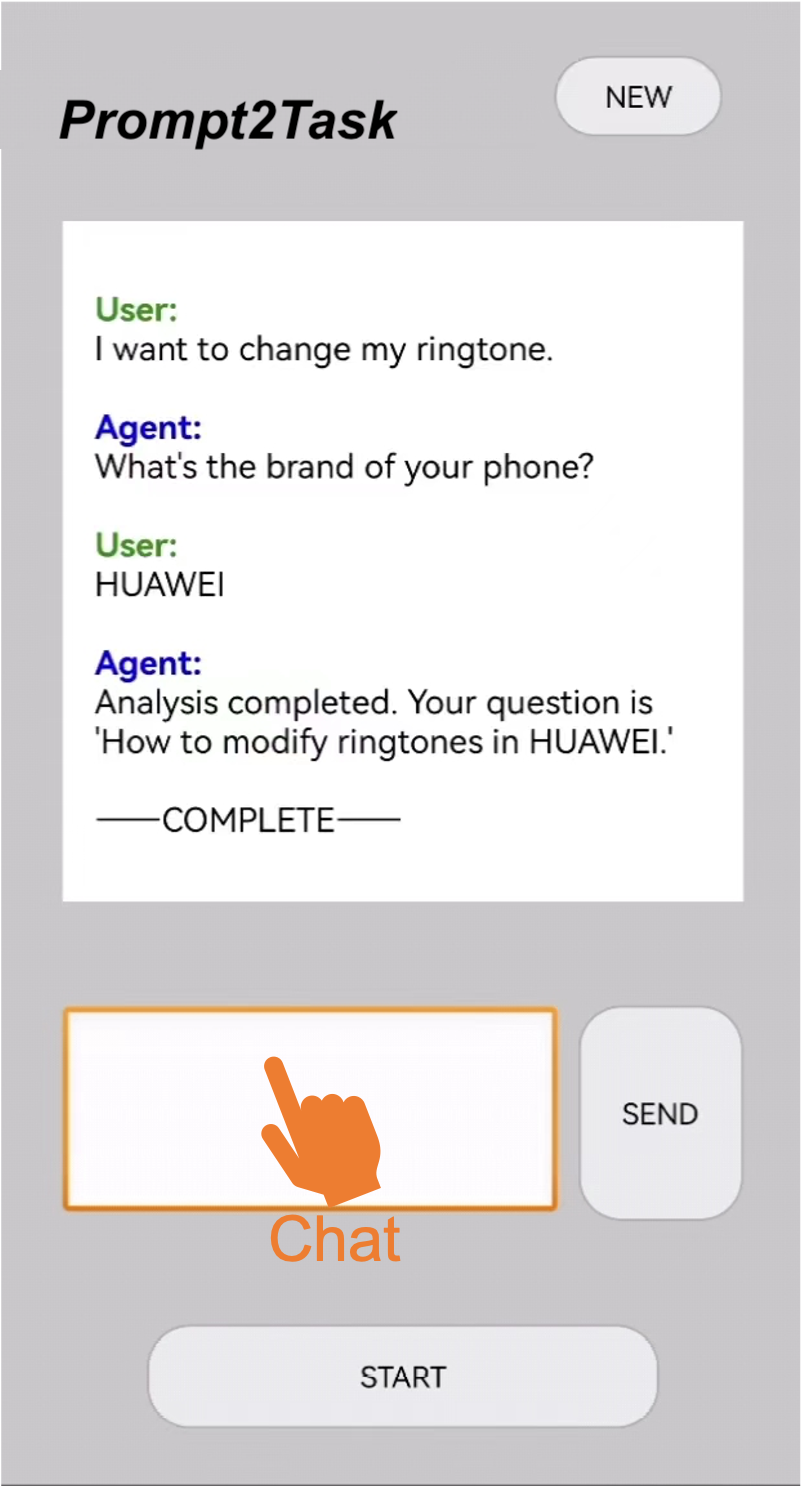}
		\end{minipage}
		\label{fig:prototype1}
	}
\subfigure[]{
		\begin{minipage}[t]{0.18\textwidth}
		\centering
			\includegraphics[height=4.6cm, keepaspectratio]{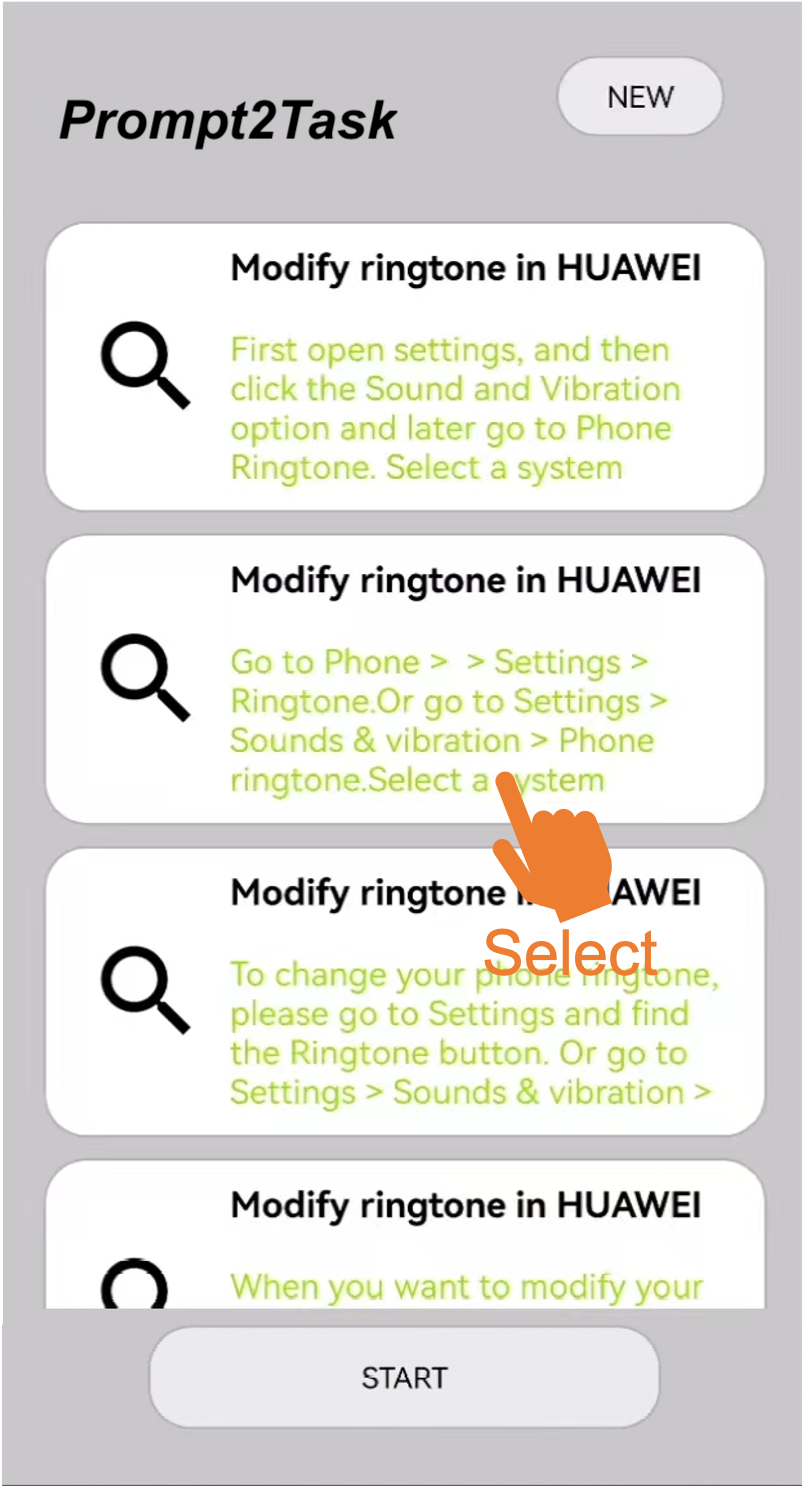}
		\end{minipage}
		\label{fig:prototype2}
	}
\subfigure[]{
		\begin{minipage}[t]{0.18\textwidth}
		\centering
			\includegraphics[height=4.6cm, keepaspectratio]{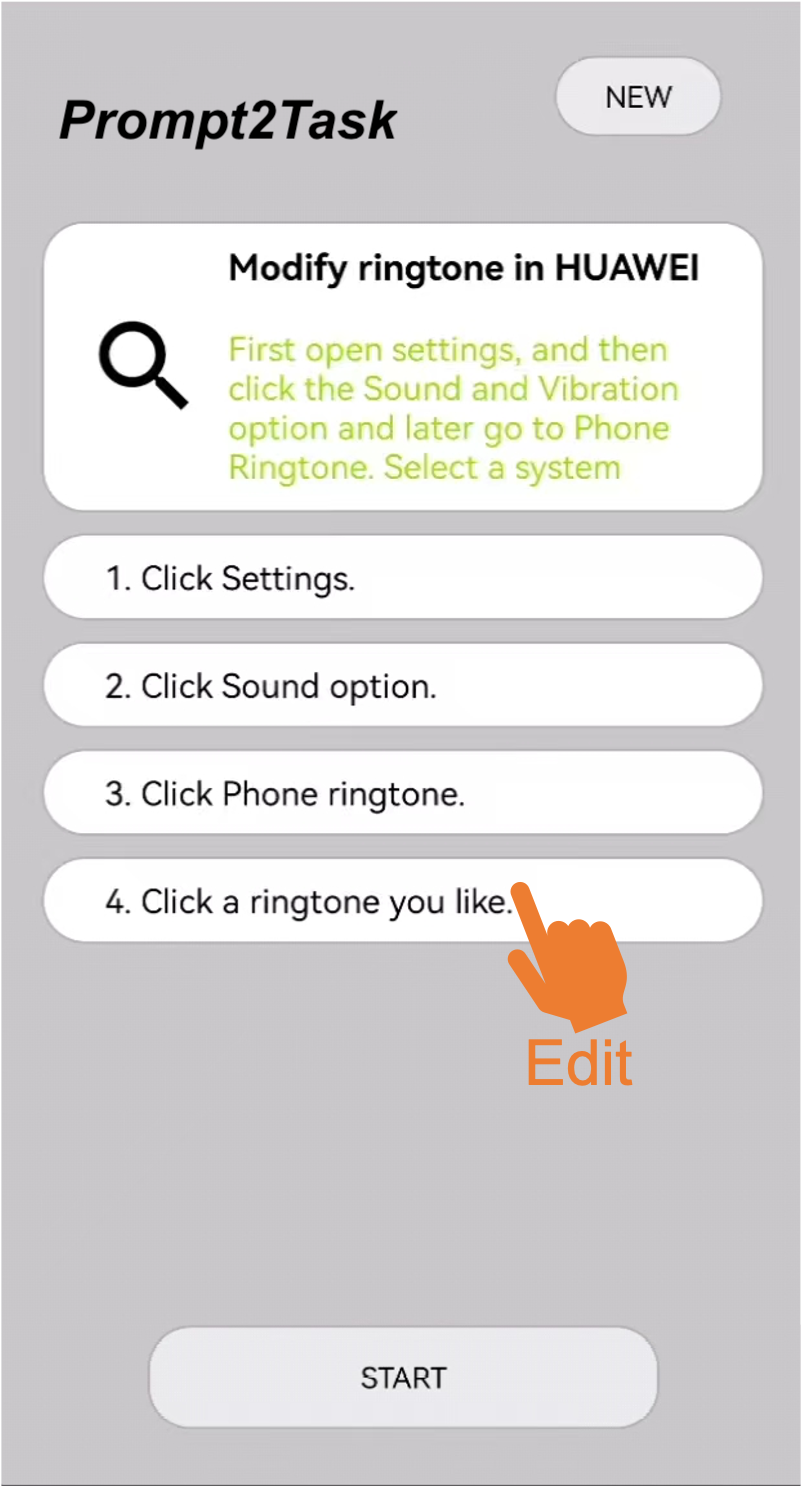}
		\end{minipage}
		\label{fig:prototype3}
	}
\subfigure[]{
		\begin{minipage}[t]{0.18\textwidth}
		\centering
			\includegraphics[height=4.6cm, keepaspectratio]{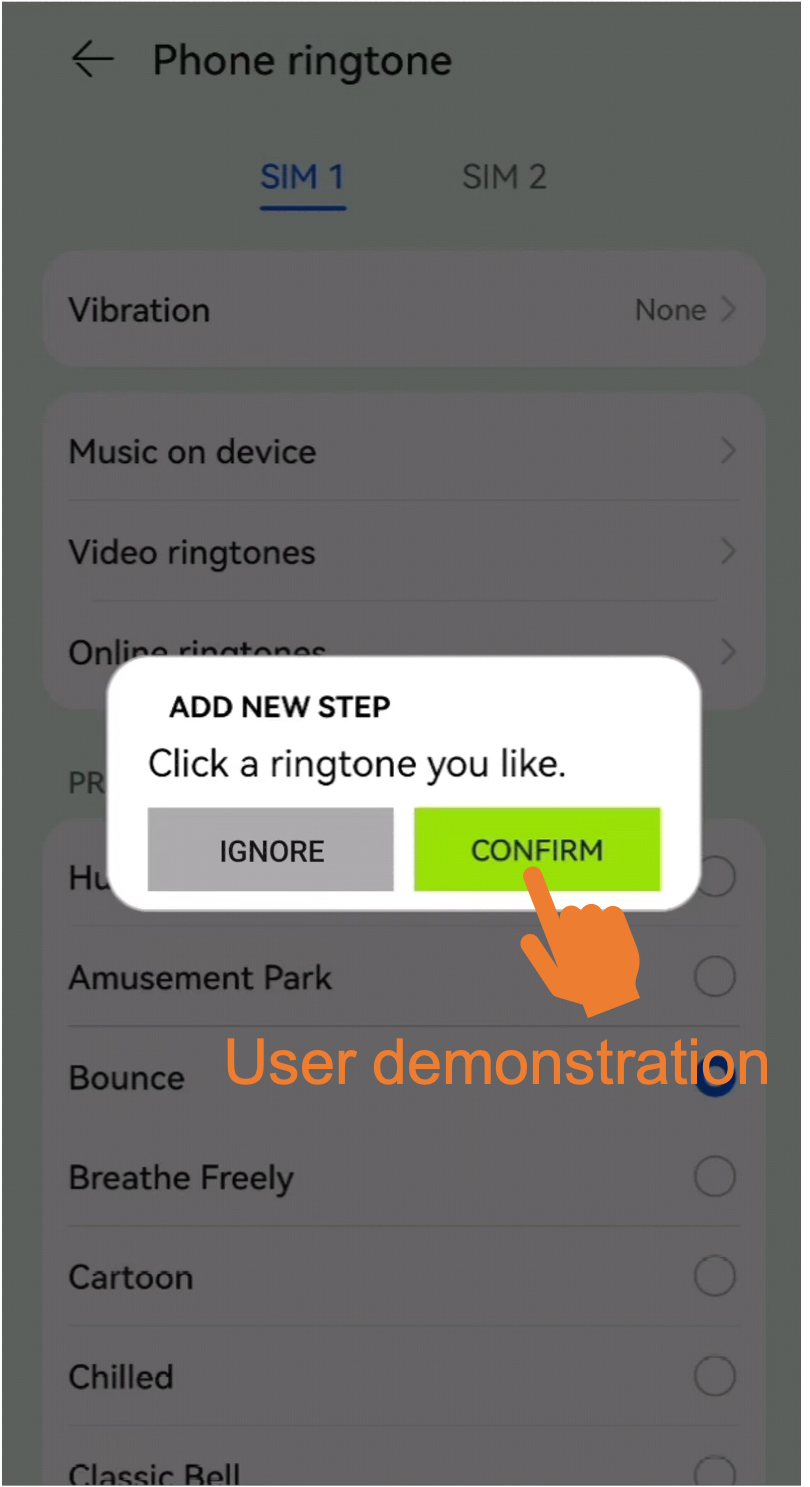}
		\end{minipage}
		\label{fig:prototype4}
	}
\subfigure[]{
		\begin{minipage}[t]{0.18\textwidth}
		\centering
			\includegraphics[height=4.6cm, keepaspectratio]{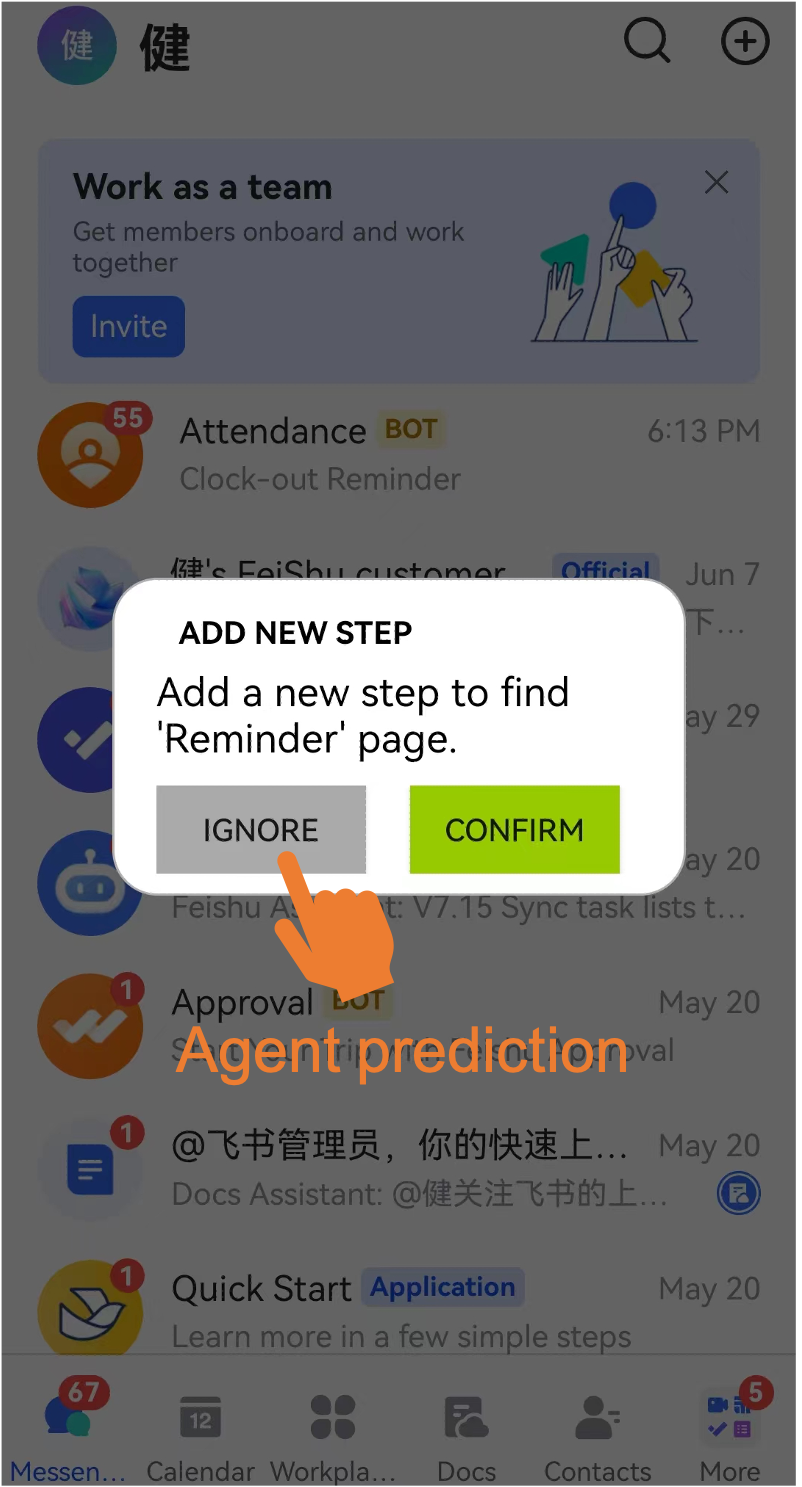}
		\end{minipage}
		\label{fig:prototype5}
	}


    \caption{\textcolor{revise}{Runtime screenshots of the Prompt2Task client. (a) The user chats with the \emph{Analysis Agent} to finalize the function description. (b) The \emph{Retrieval Agent} provides 5 tutorials for the user to choose from. (c) The \emph{Parsing Agent} generates instructions, which the user can edit. In (a) (b) (c), if the user prefers not to intervene, they can click ``START'' to skip. (d) The \emph{Grounding Agent} fails to find the target widget and requests the user to provide a demonstration. The user can begin the demonstration by selecting ``CONFIRM''. (e) The \emph{Assessment Agent} is unconfident in the current predicted operation and requests user intervention, but the user can still choose ``IGNORE'', allowing the agent to proceed with its prediction.}}
    \label{fig:prototype}
\end{figure}




\textcolor{revise}{Prompt2Task is divided into two primary components: a server and a client. The server handles information collection, instruction generation, and knowledge accumulation. It is developed using the Flask framework and crawls search results from Baidu, Bing, and Google by simulating web browser actions. The client is an Android application installed on smartphones, designed to interact with users and execute automation tasks. As illustrated in Figure \ref{fig:prototype}, the client provides a GUI for user input, selection, modification, and demonstration. During task execution, it runs in the background on the smartphone and utilizes the Accessibility Service API \footnote{\href{https://developer.android.com/reference/android/accessibilityservice/AccessibilityService}{https://developer.android.com/reference/android/accessibilityservice/AccessibilityService}} to access page layout data, and the MediaProjection API \footnote{\href{https://developer.android.com/reference/android/media/projection/MediaProjection}{https://developer.android.com/reference/android/media/projection/MediaProjection}} to capture screenshots. The client triggers operations on the smartphone through the Accessibility Service API. When user demonstration is required, an overlay is provided to monitor the user's demonstrations. Actions captured by this overlay are recorded by the system and then transferred to the native mobile GUI to trigger the corresponding operations. Prompt2Task's agents utilize the GPT-4 \footnote{\href{https://platform.openai.com/docs/models/gpt-4}{https://platform.openai.com/docs/models/gpt-4}} API for their large language model needs. All involved prompts are detailed in the appendix. In tasks involving textual similarity, including the \emph{Retrieval Agent} finding the most relevant historical task model, the \emph{Parsing Agent} finding the most similar step descriptions, and the \emph{Grounding Agent} finding the most similar instructions, we use LaBSE \cite{feng_language-agnostic_2022}, which is a state-of-the-art multilingual sentence embedding model, offers an effective trade-off between efficiency and accuracy.}


\section{Performance Evaluation}


This section assesses the performance of Prompt2Task's three-stage pipeline---information collection, instruction generation, and operation mapping---as well as the overall system performance.

\subsection{Experiment Setup}
\label{sec:pipelinesetup}
Prompt2Task is designed to fulfill a wide range of real-world needs through textual prompts provided by users, which involve various applications and built-in functionalities. Therefore, we specifically created a dataset for the experiment.


Targeting Chinese speakers, we selected 10 Chinese applications (9 applications and 1 system) for evaluation, as shown in Table \ref{tab:applicationlist}. Five of them are top-ranked in terms of active users as of 2021, and four have rankings beyond 100. The diverse selection enables us to evaluate Prompt2Task's performance across different user activity levels and application complexities.
We constructed 100 ground-truth Prompt2Task models by the following steps:
\begin{itemize}

\item \textcolor{revise}{We collected 37,987 questions related to these applications by crawling a recognized online Q\&A platform\footnote{\href{https://jingyan.baidu.com/}{https://jingyan.baidu.com/}}. The questions were sorted based on their view counts. The top half were categorized as common, while the bottom half were categorized as uncommon. We manually selected 5 common and 5 uncommon questions for each application, ensuring all $10 \times 10$ questions were unique.} Each question was used to construct one Prompt2Task model and each was manually annotated with a ground-truth function description.
\item For each model, we manually identified the highest-quality tutorial on tutorial websites based on the given function description. This tutorial text serves as the ground-truth step description for the model.
\item \textcolor{revise}{We utilized crowdsourcing to collect textual prompts that participants would use to generate automation tasks. The tasks were hosted on a public crowdsourcing website we developed, where participants were required to formulate prompts based on the ground-truth function descriptions provided to them. Each participant was instructed to ensure the uniqueness and relevance of their submissions. After filtering out duplicate, low-quality, or irrelevant prompts, we collected 25 different textual prompts for each automation task, for a total of $25 \times 100$ prompts.} Among them, 100 prompts contained complete step descriptions, without the need for further tutorial retrieval.

\item According to the step description, we then manually annotated the ground-truth instructions for each Prompt2Task model, totaling 533 instructions.
\item Finally, the experiment was conducted on an HONOR V20. For each Prompt2Task model, we manually recorded the ground-truth operation sequence on this device, totaling 543 operations. Notably, these operations do not directly correspond to the ground-truth instructions, often due to outdated tutorials. \textcolor{revise}{Details are available in our public dataset.}
\end{itemize}

\begin{table}[tb]
	\centering
	\caption{Applications and tasks overview in the experiment. WeChat, QQ, TikTok, Beautiful Weather, Safe and Sound, and Film Encyclopedia were also selected for the user evaluation in Section \ref{sec:userstudy}.}
        \label{tab:applicationlist}
	\begin{tabular}{lllll}
		\toprule  
            \hline
		\multirow{2}{*}{\textbf{Application}} & \multirow{2}{*}{\textbf{Category}} & \textbf{Activity} & \textcolor{revise}{\textbf{Step}} &\multirow{2}{*}{\textbf{Task Examples}} \\ 
		 &  & \textbf{Ranking} &\textcolor{revise}{\textbf{Range}}  & \\ 
		\Xhline{1pt}  
		\multirow{2}{*}{WeChat} & \multirow{2}{2cm}{Message} & \multirow{2}{2cm}{1} & \textcolor{revise}{\multirow{2}{2cm}{4\textasciitilde10}} & 1. View WeChat Bean details
 \\
		                           &                         &        &                 & 2. Clear cache\\
                             \hline
        \multirow{2}{*}{QQ} & \multirow{2}{2cm}{Message} & \multirow{2}{2cm}{2} & \textcolor{revise}{\multirow{2}{2cm}{3\textasciitilde8}}& 1. Share QQ screen\\
		                     &                         &  &                       & 2. Provide feedback on an issue\\
                       \hline
        \multirow{2}{*}{TikTok} & \multirow{2}{2cm}{Social Media} & \multirow{2}{2cm}{3} & \textcolor{revise}{\multirow{2}{2cm}{3\textasciitilde10}}& 1. Cancel the monthly payment \\
		                         &                               &        &                 & 2. View TikTok shop \\
                           \hline
        \multirow{2}{*}{Alipay} & \multirow{2}{2cm}{Payment} & \multirow{2}{2cm}{4} & \textcolor{revise}{\multirow{2}{2cm}{3\textasciitilde8}}& 1. Set the notification sound\\
		                         &     &                  &                       & 2. View Alipay points  \\
                           \hline
        \multirow{2}{*}{Weibo} & \multirow{2}{2cm}{Social Media} & \multirow{2}{2cm}{5}& \textcolor{revise}{\multirow{2}{2cm}{4\textasciitilde7}} & 1. Check creation earnings \\
		                        &   &                          &                         & 2. View corrected followers\\
                          \hline
                \multirow{2}{*}{Lark} & \multirow{2}{2cm}{Office} & \multirow{2}{2cm}{173} & \textcolor{revise}{\multirow{2}{2cm}{5\textasciitilde10}}& 1. Log out\\
		                              &                        & &                        & 2. Add Lark reminders to the workspace\\
                                \hline
        Beautiful & \multirow{2}{2cm}{Weather} & \multirow{2}{2cm}{366} & \textcolor{revise}{\multirow{2}{2cm}{3\textasciitilde7}} & 1. Update the weather every 3 hours\\
	Weather	                                    &                        &    &                     & 2. Observe the weather in Beijing\\
                                      \hline
        Safe and  & \multirow{2}{2cm}{Lifestyle} & \multirow{2}{2cm}{392}& \textcolor{revise}{\multirow{2}{2cm}{3\textasciitilde8}} & 1. Set the shipping address
\\
		     Sound                            &             &               &                         & 2. Enable the night mode\\
                                   \hline
        Film& \multirow{2}{2cm}{Entertainment} & \multirow{2}{*}{671}& \textcolor{revise}{\multirow{2}{*}{4\textasciitilde10}} & 1. Join the user experience program
\\
	Encyclopedia	                                       &                               &                    &     & 2. Bind QQ with the Film Encyclopedia  \\
                                         \hline
        \multirow{2}{*}{System Tools}   & \multirow{2}{2cm}{System} & \multirow{2}{2cm}{-}& \textcolor{revise}{\multirow{2}{2cm}{4\textasciitilde8}} & 1. Turn off the pure mode\\
		                                         &                       &    &                   & 2. Hide photos  \\
		\bottomrule  
	\end{tabular}
\end{table}

\subsection{Experiment Method}
\label{sec:experimentmethod}


In our experiment, we evaluated three configurations of Prompt2Task:
\begin{itemize}
    \item Standard Prompt2Task.
    \item \textbf{Prompt2Task(AK)}: Prompt2Task enhanced with accumulated knowledge.
    \item \textcolor{toconfirm}{\textbf{Prompt2Task(AK+minimal UI)}: Prompt2Task enhanced with accumulated knowledge and minimal necessary user intervention. Only essential user interventions during the \textbf{operation mapping} stage to provide important parameters are permitted. For instance, ``Click a ringtone you like'' in Figure \ref{fig:title} does require user intervention, while the scenario in Figure \ref{fig:demonstrate} does not inherently require it but arises due to the agent's limitations.}
    \item \textcolor{toconfirm}{\textbf{Prompt2Task(AK+UI)}: Prompt2Task further enhanced with both accumulated knowledge and user intervention, providing perfect user intervention to maximize success rate.}
\end{itemize}


\begin{sloppypar}
The experiment assessed Prompt2Task's performance at individual stages and across the entire pipeline. Notably, in a single stage where a task was not fully completed, without user intervention, the knowledge that Prompt2Task could accumulate was quite limited. Therefore, \textbf{Prompt2Task(AK)} was evaluated solely on its entire pipeline performance. When evaluating the performance of individual stages, we used ground-truth inputs, ensuring that any output deviation stemmed exclusively from the current stage, eliminating errors from preceding stages. Specifically, in the instruction generation stage, we input 100 ground-truth function and step descriptions; for the operation mapping stage, we input 100 ground-truth instruction lists. In evaluating the information collection stage and the entire pipeline, the input comprised 2,500 original textual prompts. 
To minimize the impact of input order and the inherent variability of LLM outputs, we randomly shuffled the inputs and conducted five rounds of testing to calculate the average performance. 
\end{sloppypar}


We selected appropriate baselines for comparison at both individual stages and the entire process. For the information collection stage, manual queries on Google, Bing, and Baidu served as the baseline. For instruction generation, we used Seq2act \cite{li_mapping_2020} as the baseline. For the operation mapping stage, we employed strict text matching as the baseline. In assessing the entire pipeline, our baseline was using GPT-4 to generate instructions, followed by strict text matching for operation mapping. Additionally, considering some works that use step descriptions as input, we used LLM4Mobile \cite{wang_enabling_2023} and AdbGPT \cite{feng2023prompting} as baselines for the last two stages combined.

\subsection{Experiment Results}

The experiment results are organized based on the three stages within Prompt2Task as well as the performance of the entire pipeline.


\subsubsection{Information Collection}



This stage focuses on evaluating whether Prompt2Task can successfully generate a function description and step description based on a user-provided textual prompt. 


Among the 2,500 textual prompts, 100 contained complete step information and didn't require further online retrieval. Users accepted the analyzed results for 97.00\% of these prompts, with a few exceptions due to the LLM's data fabrication issues.
The remaining 2,400 prompts were used to assess the top 5 tutorials retrieved by the system. The criteria for a ``hit'' was whether users agreed that a tutorial matched the ground-truth function description, regardless of on-phone executability. A tutorial at the top indicated that the system could automatically select it, eliminating the need for user intervention. \textcolor{revise}{Among the 2,400 prompts, 12.21\% have incomplete descriptions of functions and required further supplementation.}

\textcolor{toconfirm}{Before the experiment, we benchmarked the search capabilities of Google, Bing, and Baidu using 100 ground-truth function descriptions. As Table \ref{tab:searchengine} shows, Baidu outperformed the others, achieving a 100.00\% Top-5 hit rate and a 99.00\% Top-1 rate. The results demonstrate that all tasks could be matched with online tutorials, provided the descriptions are sufficiently standard and accurate.}

\begin{table}[t]
	\centering
	\caption{Performance of various search engines in retrieving tutorials, using ground-truth function descriptions as input.}
        \label{tab:searchengine}
	\begin{tabular}{lccc}
		\toprule  
		&Top-5&Average&Top-1\\
            &Hit Rate(\%)&Rank&Hit Rate(\%)\\
		\midrule  
            Google &93.00&1.23 (sd=0.59)&79.00\\
            Bing &96.00&1.25 (sd=0.70)&81.00\\
		Baidu &100.00&1.01 (sd=0.10)&99.00\\
		\bottomrule  
	\end{tabular}
\end{table}

\begin{table}[t]
	\centering
	\caption{Performance of information collection, \textcolor{revise}{using 2,400 textual prompts as input.}}
        \label{tab:searching}
	\begin{tabular}{l|ccc|c}
		\toprule  
            &\multicolumn{3}{c|}{\textbf{Tutorials Match Function Description}}&\multicolumn{1}{c}{\textbf{Step Description}}\\
		\cline{2-5}  
		&Top-5&Average&Top-1&Accuracy\\
            &Hit Rate(\%)&Rank&Hit Rate(\%)&(\%)\\
		\hline  
		Manual(Google)&86.71&1.55 (sd=0.85)&64.92&-\\
		Manual(Bing)&89.17&1.36 (sd=0.80)&69.38&-\\
		Manual(Baidu)&97.21&1.42 (sd=0.76)&87.96&-\\
            Prompt2Task&98.83&1.27 (sd=0.91)&\textbf{89.38}&\textbf{88.38}\\
            Prompt2Task(AK+UI)&100.00&1.22 (sd=0.74)&\textbf{90.83}&\textbf{99.21}\\
		\bottomrule  
	\end{tabular}
\end{table}



However, prompts provided by users often lack accuracy, leading to a decrease in effectiveness when using the original prompts for manual searches on these websites. Using them as baselines, we tested the performance of Prompt2Task. As shown in Table \ref{tab:searching}, Prompt2Task improved the Top-5 hit rate from 97.21\% (Baidu) to 100\%, while achieving a Top-1 hit rate of 90.83\%. \textcolor{revise}{On average, each prompt required 0.15 user interventions (sd = 0.52), which represents the number of dialogue turns between the user and Prompt2Task (excluding the initial input).} Remarkably, without user interventions, Prompt2Task demonstrated a Top-1 hit rate of 89.38\%, surpassing the baseline performance.

After the agent matches function descriptions with online tutorials, it extracts concise step descriptions from complex web data. The accuracy of step descriptions refers to the correct extraction from web pages, ensuring they are free from fabrication, omissions, or errors. When Prompt2Task automatically selected the first candidate, the accuracy was 88.38\%. When users could select from five candidates, the accuracy rose to 99.21\%.


\subsubsection{Instruction Generation}

This stage evaluates Prompt2Task's ability to transform step descriptions into a list of instructions in the predefined format. 

Using the ground-truth step descriptions of the selected 100 Prompt2Task models as input, we evaluated correctness using two metrics, focusing solely on textual interpretation:

\begin{itemize}
\item \textbf{Complete Accuracy}: The percentage of Prompt2Task models whose all generated instructions match the ground truth.
\item \textbf{Partial Accuracy}: The proportion of correct instructions relative to the total ground-truth instructions.
\end{itemize}

We used the Seq2Act model \cite{li_mapping_2020}, a transformer trained using the ANDROIDHOWTO dataset, as our baseline (after translating
step descriptions into English). Table \ref{tab:parsing} reveals that Prompt2Task achieved 100\% accuracy for both complete and partial metrics. On average, achieving perfection required 0.93 user interventions per task (sd=1.34). It's worth noting that while perfection necessitated user interventions, the subsequent pipeline stages are designed to accommodate minor errors, reducing the need for strict user oversight. Without user intervention, Prompt2Task attained 84.54\% partial and 73.00\% complete accuracy. In cold-start scenarios, Prompt2Task has achieved performance comparable to that achieved by related works \cite{li_mapping_2020, 10.1145/3581998}, which require substantial training data (at least 10K-level).

\begin{table}[t]
	\centering
	\caption{Performance of instruction generation.}
        \label{tab:parsing}
	\begin{tabular}{lccc}
		\toprule  
		&Partial Accuracy(\%)&Complete Accuracy(\%)&User Intervention\\
		\midrule  
		Seq2act \cite{li_mapping_2020}&70.13&54.00&-\\
            Prompt2Task&84.54&\textbf{73.00}&-\\
            Prompt2Task(AK+UI)&100.00&\textbf{100.00}&0.93 (sd = 1.34)\\
		\bottomrule  
	\end{tabular}
\end{table}

\subsubsection{Operation Mapping}

\begin{sloppypar}
This stage evaluates Prompt2Task's ability to execute operations on smartphones, guided by instructions. We used the ground-truth instruction lists of the 100 Prompt2Task models as input.
\end{sloppypar}

Our baseline for comparison utilized strict text matching to identify operation widgets. Table \ref{tab:grounding} shows the results. With user intervention (i.e., users can demonstrate any operations), the success rate reached 96.00\%. The remaining 4\% of tasks were unfeasible under the current test environment. On average, each task required 0.68 user interventions (sd = 0.85). Even without user assistance, the success rate was 62.00\%, far surpassing the baseline. When users were allowed to provide necessary parameters, the success rate was 91.00\%, indicating that 29.00\% of the tasks require user-provided information even under optimal conditions.


\textcolor{revise}{In terms of the detailed performance without user intervention, Table \ref{tab:groundingstrategy} presents the frequency and accuracy of the \emph{Grounding Agent}'s strategies when handling new instructions. The agent seldom uses the ``expand'' strategy, which also shows the lowest accuracy. This indicates that even though some instructions are wrong or outdated, they still hold great reference value. In most cases, the agent prefers to use ``redirect'' or ``add'' strategies to adapt to the current operating environment. Table \ref{tab:assessmentstrategy} shows the frequency and accuracy of the \emph{Assessment Agent}'s strategies for the operations predicted by the \emph{Grounding Agent}. The agent lacked confidence in 16.53\% of the decisions, of which 68.57\% required necessary user intervention to complete the task. In the remaining cases,
while user intervention is optional, it can help the agent make more informed decisions.}

\begin{table}[t]
	\centering
        \caption{\textcolor{revise}{Performance of operation mapping. Prediction accuracy is measured as the proportion of correctly executed operations out of the total number of operations, where an operation is considered correct if it is appropriate in the context at that time. The success rate is determined by the ratio of completed tasks.}}
        \label{tab:grounding}
	\begin{tabular}{lccc}
		\toprule   
		&Prediction& Success&User\\
		&Accuracy(\%) & Rate(\%)&Intervention\\
		\midrule  
		Text only&61.49&44.00&-\\
            Prompt2Task&80.25&\textbf{62.00}&-\\
            Prompt2Task(AK+minimal UI)&88.06&\textbf{91.00}&0.61 (sd = 0.77)\\
            Prompt2Task(AK+UI)&92.11&\textbf{96.00}&0.68 (sd = 0.85)\\
		\bottomrule  
	\end{tabular}
\end{table}

\begin{table}[tb]
	\centering
        \caption{\textcolor{revise}{Frequency and accuracy of the \emph{Grounding Agent}'s strategies. The ``block'' strategy indicates the agent could not predict the operation, thus it has no accuracy measure.}}
        \label{tab:groundingstrategy}
	\begin{tabular}{lccc}
		\toprule   
		&Frequency(\%)&Accuracy(\%)\\
		\midrule  
            Strict Text Matching &26.44& 100.00\\
            Redirect &30.17&90.72\\
            Add &39.50&56.69\\
            Skip &1.24&100.00 \\
            Expand &1.40& 33.33\\
            Block &1.24&- \\
		\bottomrule  
	\end{tabular}
\end{table}

\begin{table}[tb]
	\centering
        \caption{\textcolor{revise}{Frequency and accuracy of the \emph{Assessment Agent}'s strategies.}} 
        \label{tab:assessmentstrategy}
	\begin{tabular}{lccc}
		\toprule   
		&Frequency(\%)&Accuracy(\%)\\
		\midrule  
            Follow &81.89&83.46\\
            Change &5.04&53.13\\
            Retract &3.31&95.24\\
            Finish &9.76&100.00\\
		\bottomrule  
	\end{tabular}
\end{table}

\subsubsection{The Entire Pipeline}

To evaluate the overall effectiveness of Prompt2Task, we input 2,500 textual prompts into the pipeline, which comprises \textbf{information collection}, \textbf{instruction generation}, and \textbf{operation mapping}. 

\textcolor{revise}{Given the task-related textual inputs, we did not use UGIF \cite {venkatesh_ugif_2022} and CogAgent \cite{hong2023cogagent} as baselines, as it would be unreasonable to require them to support our tasks without pre-provided tutorial sets or fine-tuning data. Additionally, we did not use Voicify \cite{10.1145/3581998}, as our verification showed that only 15\% of the tasks' target pages had intent filters configured in the app code, which means that it is impossible to reach the remaining task pages using only page descriptions.} Therefore, we established our baseline using GPT-4 for direct instruction generation from user inputs, coupled with strict text matching for operations.

As shown in Table \ref{tab:overall}, the success rate for each stage is strictly determined by the complete accuracy achieved up to that stage. Importantly, the overall success rate is not a simple product of success rates at each stage, due to the system's built-in flexibility for error correction in downstream stages. For instance, the operation mapping stage doesn't strictly follow the generated instructions; it also utilizes the agent's exploratory strategies. \textcolor{revise}{For \textbf{Prompt2Task(AK+minimal UI)}, the average number of user interventions per task is 0.42 (sd = 1.32), excluding the initial user input. For \textbf{Prompt2Task(AK+UI)}, the average number is 0.69 (sd = 1.31).} Notably, this number is smaller than the sum of user interventions required for each stage as reported earlier, which shows that the accumulation of knowledge from completed tasks can reduce user interventions.



\textcolor{minorrevise}{Additionally, we compared Prompt2Task with baselines that support only step description inputs. We tested the performance using ground-truth step descriptions as inputs, which eliminates errors in online tutorial retrieval. Notably, it does not guarantee that the ground-truth tutorial is entirely feasible in the current test environment. While 31\% of the tasks involve outdated step descriptions, only 11.82\% of the individual steps are outdated, which means that a significant portion of the steps are still correct, and only minor corrections may be needed to complete the tasks. As shown in Table \ref{tab:twostages}, Prompt2Task handles outdated tutorials effectively, achieving a 96\% success rate when user intervention is applied. The discussion on outdated tutorials is detailed in Section \ref{sec:knowledgediscussion}. Moreover, the prediction accuracy of \textbf{Prompt2Task(AK+minimal UI)} and \textbf{Prompt2Task(AK+UI)} is lower than the success rate because the system executed redundant exploratory steps, which also led to a decrease in relative efficiency. However, for the goal of fully completing the task, this trade-off is considered acceptable.}


\begin{table}[t]
	\centering
  \caption{Success rate (\%) across various stages, using 2,500 textual prompts as input.} 
        \label{tab:overall}
	\begin{tabular}{lccc}
		\toprule  
		& Information Collection & Instruction Generation & Operation Mapping\\ 
		\midrule  
		GPT-4 only, Text only&-& 50.88 &22.28\\
            Prompt2Task&88.42& 64.92 &\textbf{60.80}\\
            Prompt2Task(AK)&92.26& 72.32 & \textbf{78.16}\\
            Prompt2Task(AK+minimal UI)&93.16& 73.36&\textbf{90.68}\\
            Prompt2Task(AK+UI)&99.12& 99.12 & \textbf{95.24}\\
		\bottomrule  
	\end{tabular}
\end{table}

\begin{table}[t]
	\centering
  \caption{\textcolor{revise}{The performance of the entire system with ground-truth step descriptions as input. The definitions of prediction accuracy and success rate are the same as those in Table \ref{tab:grounding}. Relative efficiency is the ratio of ground-truth operations to actual operations performed.}} 
        \label{tab:twostages}
	\begin{tabular}{lccc}
		\toprule  
		   & Prediction Accuracy(\%)  & Success Rate(\%)& Relative Efficiency(\%)\\ 
		\midrule  
            LLM4Mobile \cite{wang_enabling_2023}&54.58&23.00&89.68\\
            AdbGPT \cite{feng2023prompting}&65.97&36.00&87.86\\
            Prompt2Task&80.15&\textbf{62.00}&85.03\\
            Prompt2Task(AK+minimal UI)&87.78&\textbf{91.00}&76.23\\
            Prompt2Task(AK+UI)&91.46&\textbf{96.00}&76.13\\
		\bottomrule  
	\end{tabular}
\end{table}

\subsection{Error Analysis}
We analyzed the errors that occurred in the three stages of Prompt2Task.

\subsubsection{Information Collection}

Prompt2Task effectively matches tutorial web pages but falls short in content extraction. Two main issues affect the performance. First, web pages containing excessively long text compromise the LLM's extraction capabilities, leading to information loss and increased risk of generating inaccurate content. Second, the LLM also tends to generate fabricated step descriptions when interpreting the textual prompts.

To address these challenges, we have considered two options for improving text quality. Initially, we considered implementing a specialized assessment agent, similar to the one used in the \textbf{operation mapping}, to manage text quality. However, we ultimately chose a human-in-the-loop approach, considering the efficiency of manual text assessment and the extra time needed for an agent-based solution. Our current system allows users to manually select high-quality step descriptions for subsequent stages.

\subsubsection{Instruction Generation}

\begin{figure}[tb]
	\centering
	\begin{minipage}{0.48\linewidth}
		\centering
		\includegraphics[height=4.4cm, keepaspectratio]{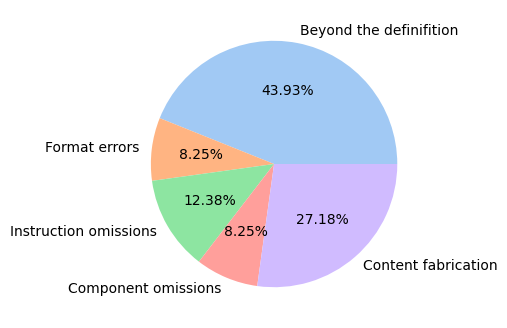}
		\caption{Errors of instruction generation.}
		\label{fig:parsingerror}
	\end{minipage}
	\begin{minipage}{0.48\linewidth}
		\centering
		\includegraphics[height=4.4cm, keepaspectratio]{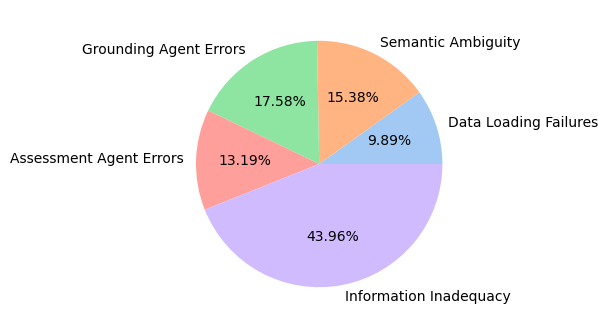}
		\caption{Errors of operation mapping.}
		\label{fig:groundingerror}
	\end{minipage}
\end{figure}

Prompt2Task's automatic instruction generation achieves a partial accuracy of 84.54\%, with errors distributed across six main categories as shown in Figure \ref{fig:parsingerror}. These errors are generally resolved through simple user corrections. Specifically, 43.93\% of errors are due to instructions that go beyond the existing definition (Table \ref{tab:operation}). Within this subset, 24.51\% involve the LLM using near-synonyms (e.g., ``drag down'' instead of ``scroll down''). The remaining 19.42\% pertain to complex operations like ``log in'' or ``register'' that the LLM struggles to parse, which can be potentially resolved by the exploration in the subsequent stage. Additionally, ``Content fabrication'' represents the second largest category of errors, which is a notable issue of the LLM. It occasionally does not strictly adhere to the given step description and instead invents or elaborates on instructions.

Notably, not all errors necessitate user intervention. Many issues are addressed in the subsequent stage, improving the overall success rate compared to the instruction generation stage (Table \ref{tab:overall}). It is due to the system's ability to self-correct certain operations. Essentially, generated instructions act as guidelines rather than absolutes, and even incorrect instructions offer valuable context that enhances the system's overall performance.

\subsubsection{Operation Mapping}



Prompt2Task achieves an 88.06\% prediction accuracy with minimal necessary user interventions. As detailed in Figure \ref{fig:groundingerror}, errors can be categorized as follows:

\textbf{Data Loading Failures}: 9.89\% are caused by issues in page data loading, making the system fail to access GUI widgets.
  
\textbf{Semantic Ambiguity}: 15.38\% of errors arise from the system misidentifying widgets due to semantic ambiguities and fuzzy matching challenges. In this case, the agent can iterate attempts and refine the operation sequence based on triggered results.

\textbf{Agent-Related Errors}: 17.58\% of errors are due to the \emph{Grounding Agent} taking an unrecoverable erroneous direction in task execution. 13.19\% of errors happen when the \emph{Assessment Agent} wrongly rejects operations initially predicted by the \emph{Grounding Agent}. Such errors occur when contextual information is insufficient, but they diminish as the \emph{mobile interaction graph} expands. The graph can provide the agents with a richer context for each operation's implications.

\textbf{Information Inadequacy}: Errors due to incomplete information constitute the largest proportion, accounting for 43.96\%. Widgets without textual descriptions are particularly affected. Although the mobile semantics module in the \emph{Grounding Agent} employs advanced image-to-text models, their ability to recognize the fine-grained semantics of widgets is limited. Fortunately, users can help interpret the widgets by demonstrating how the instructions map to the corresponding widgets. User demonstration can also provide context-specific interpretations, such as clarifying that ``select'' in a certain context means ``switch'' rather than ``click''.

\section{User Evaluation}
\label{sec:userstudy}









Given the influence of user intervention on system performance, we assessed Prompt2Task's usability through engagement with diverse users.

\subsection{Experiment Setup}


 \textbf{Participant Demographics}:
\textcolor{revise}{We recruited a total of 12 participants for our study through online postings in local communities. Each participant was compensated at a rate of \$15 per hour for their contribution. Before the study began, each participant was required to complete an informed consent form that detailed the nature of the study, the extent of data collection, and the installation of the Prompt2Task client on their personal devices (detailed in Section \ref{sec:implementation}). We confirmed that no personal data would be retained post-study without explicit permission. This study has been reviewed and approved by the appropriate Institutional Review Board (IRB).} 

The participant's proficiency was determined through a pre-study questionnaire that evaluated their comfort level and frequency with smartphone usage. The participants were divided into the following two categories according to their proficiency.

  \begin{itemize}
    \item \textit{Skilled Users}: 6 participants were proficient smartphone users (2 females; aged from 18 to 40, M = 28.83), who were familiar with a wide range of applications and functions. 
  
    \item \textit{Unskilled Users}: 6 participants were less proficient in smartphone usage (4 females; aged from 50 to 69, M = 60.00). They are elders, beginners, or infrequent smartphone users. 
  \end{itemize}

\textbf{App Selection}: 
The study included tasks across 6 applications---3 commonly used (WeChat, QQ, TikTok) and 3 less commonly used (Beautiful Weather, Safe and Sound, Film Encyclopedia). The selection criteria for these applications are consistent with those described in Section \ref{sec:pipelinesetup}.

\textbf{Task Selection}: 
To ensure balanced task assignments, the participants were grouped into 3 sets, each containing 4 participants (2 skilled and 2 unskilled users). There are two types of tasks.
  \begin{itemize}
    \item \textit{Predefined Tasks}: 
    The 4 participants within a set are assigned the same 8 predefined tasks, split across 2 applications: one commonly used and one less commonly used. For each application, 2 common and 2 uncommon automation tasks are chosen, based on criteria detailed in Section \ref{sec:pipelinesetup}. Part of the tasks are shown in Table \ref{tab:applicationlist}.
    In each four-participant set, we assume that `a' and `b' are skilled users, whereas `c' and `d' are unskilled users. To ensure balance, for each application, one common task is performed manually by `a' and `c', while `b' and `d' use Prompt2Task. For the other common task, roles are reversed: `a' and `c' use Prompt2Task, while `b' and `d' perform it manually. The approach is similar for uncommon tasks.
    
   \item \textit{User-Suggested Tasks}:
In addition to the predefined tasks, each participant is allowed to propose 2 automation tasks for each of their assigned applications. We encouraged participants to suggest challenging tasks. For these tasks, participants will employ both manual execution and Prompt2Task. To counterbalance the effects of the method sequence, the order of methods is alternated across tasks.

  \end{itemize}
  
\textbf{Methods}: Almost all tasks in the experiment, requiring in-depth application functionality, have been validated as unsupported by Google Assistant and HUAWEI Celia, which excel in information gathering but fail to execute tasks. For manual execution, participants are allowed to seek assistance from online resources and people around them. For tasks carried out using Prompt2Task, participants can freely provide any information through textual prompts, which can either describe the task's goal or outline the detailed steps required for the task. After completing the task, participants are required to use a different expression to invoke the task again. 

\textbf{Metrics}:
We recorded the time and success rate for each task, both when performed manually and via Prompt2Task. For tasks carried out using Prompt2Task, the frequency of user interventions was also recorded.

 \textbf{Experimental Device}:
Participants used their own Android smartphones for the experiment. Prompt2Task's client is installed on these devices \textcolor{revise}{(detailed in Section \ref{sec:implementation})}.

In summary,  we collected 192 ($12 \times 8 \times 2$) data points for tasks completed using Prompt2Task (half of which from first-time usage and the other half from second-time usage), and 96 ($12 \times 8$) data points for tasks completed manually.

\subsection{Experimental Results}

This section demonstrates the experiment's results, focusing on metrics including success rate, completion time, and frequency of user interventions, as well as subjective feedback.

\begin{table}[htbp]
	\centering
	\caption{The task success rate (\%) of different methods and participant groups. \textcolor{revise}{Tasks are considered successful only if completed within 10 minutes. ``Prompt2Task (Second Time)'' refers to tasks that have been previously executed and are being re-executed with a different textual prompt.}}
        \label{tab:successmethod}
	\begin{tabular}{cccc}
		\toprule   
		&Manual&Prompt2Task (First Time) & Prompt2Task (Second Time)\\
		\hline  
		Unskilled&66.67& 97.91& 93.75\\
            Skilled&77.08& 97.91& 97.91\\
		\bottomrule  
	\end{tabular}
\end{table}

\begin{figure}[htbp]
	\centering
	\begin{minipage}[t]{0.48\linewidth}
		\centering
		\includegraphics[height=5cm,keepaspectratio]{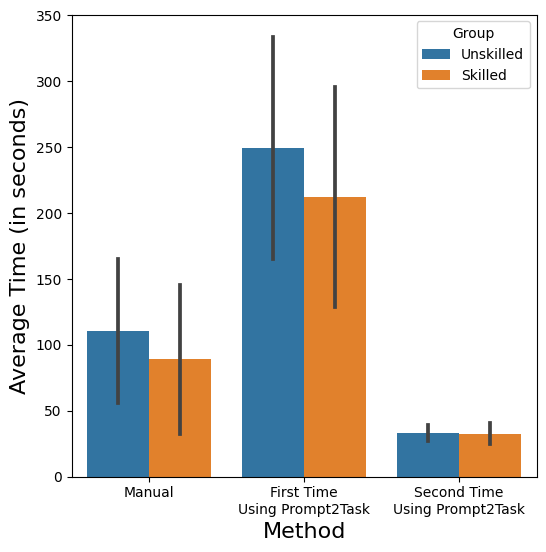}
		\caption{Impact of methods and participant groups on task completion time (detailed in Table \ref{tab:timemethod}). }
		\label{fig:timemethod}
	\end{minipage}
	\begin{minipage}[t]{0.48\linewidth}
		\centering
		\includegraphics[height=5cm,keepaspectratio]{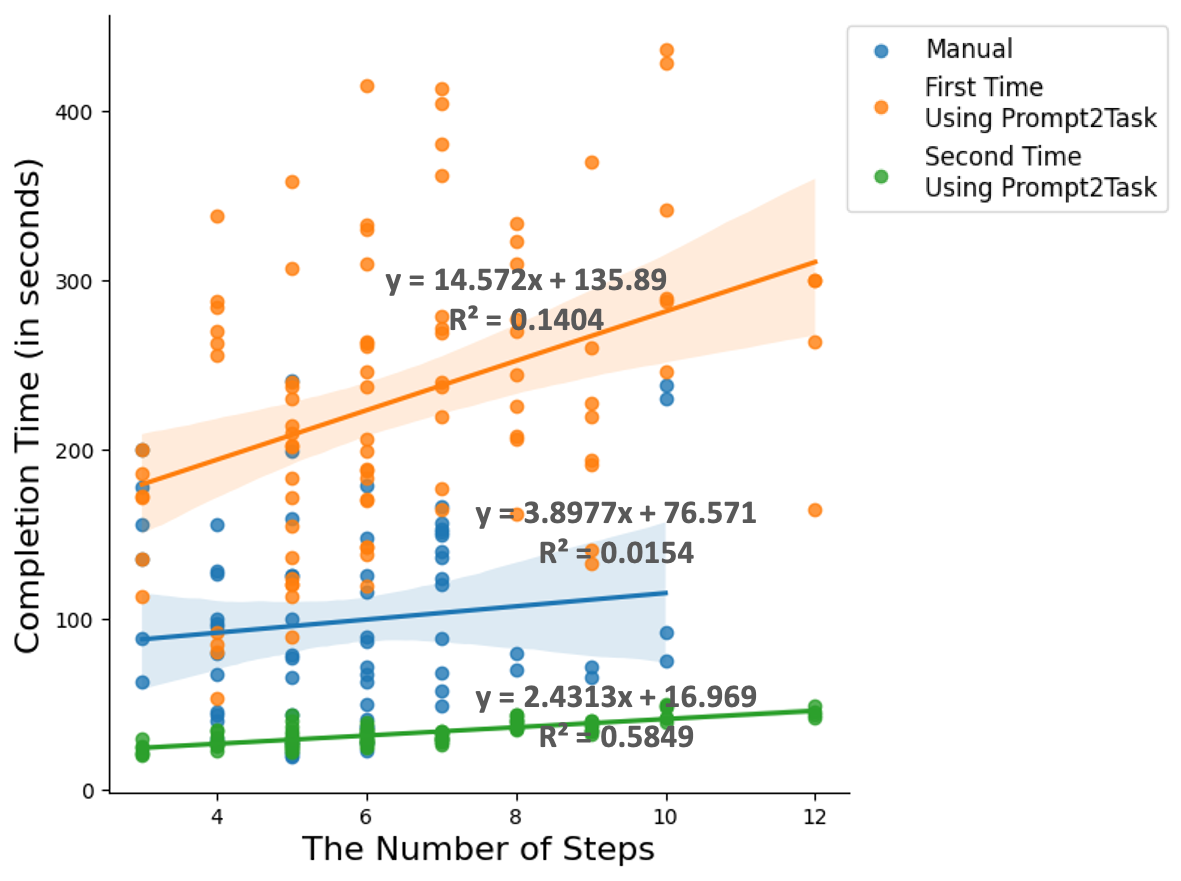}
		\caption{Impact of step count on task completion time.}
		\label{fig:timetimes}
	\end{minipage}
\end{figure}


\textbf{Success Rate.} Table \ref{tab:successmethod} reveals that Prompt2Task outperforms manual methods in both skilled and unskilled groups in terms of task success rate. Failures during the first use of Prompt2Task were largely attributable to the low quality of online tutorials for uncommonly used applications. Failures during the second attempt were mainly due to page data loading issues that blocked access to operation widgets.


\textbf{Completion Time.} Figure \ref{fig:timemethod} shows that completing tasks with Prompt2Task takes significantly longer during the first use than when carried out manually (p-value\textless 0.001). Figure \ref{fig:timetimes} indicates that the completion time correlates positively with the number of steps involved during the first use of Prompt2Task, which is more pronounced than with manual methods and the second use of Prompt2Task. The time stems primarily from the system's retrieval for online tutorials and exploration of operation sequences, which is tied to the efficiency limitations of the LLM. Notably, time decreases substantially on second use, as the system's historical task repository eliminates the need for additional tutorial retrieval and exploration, thereby becoming significantly faster than manual methods (p-value\textless 0.001).


\textbf{User Intervention.} When using Prompt2Task, each new task required 0.92 user interventions (sd = 1.07) on average. Of these, 0.75 interventions per task (sd = 1.04) occurred during the \textbf{operation mapping}. Among the user intervention requests made by the \emph{Assessment Agent}, 66.67\% were deemed necessary. Notably, the second time a task was executed using Prompt2Task, user interventions were no longer required.

\begin{figure}[tb]
  \centering
  \includegraphics[width=.75\linewidth]{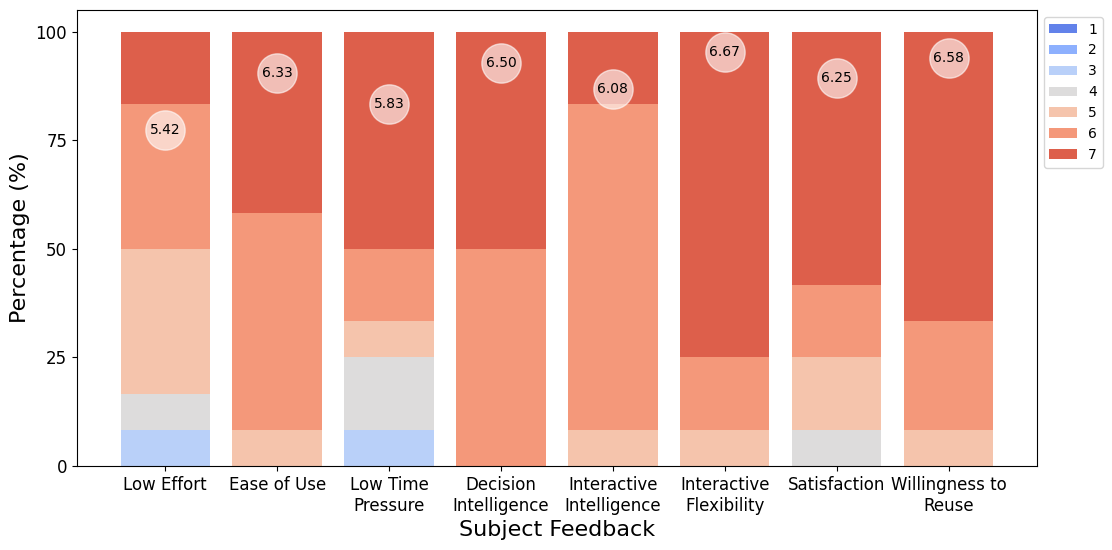}
  \caption{Distribution of participant ratings for each statement in Table \ref{tab:statement}, assessing various aspects of Prompt2Task. Ratings range from 1 (strongly disagree) to 7 (strongly agree).}
  \label{fig:statement}
\end{figure}

\textcolor{minorrevise}{
\textbf{Subjective Feedback.} Post-experiment interviews were conducted with participants, and their 7-point Likert scale responses are summarized in Figure \ref{fig:statement}. The participants' ratings demonstrate that Prompt2Task was well-received across several dimensions, including low effort, ease of use, low time pressure, decision intelligence, interactive intelligence, interactive flexibility, satisfaction, and willingness to reuse. Additionally, the following insights were gathered from the interviews:
}

\begin{itemize}

\item \textcolor{minorrevise}{\textbf{Preference for Forms of Intervention:} Participants had different experiences with various forms of intervention and expressed distinct expectations for improvement.}

\textcolor{minorrevise}{
For the dialogue, most participants (11/12) were satisfied, finding it an effective way to express their intentions. P3 expressed surprise and satisfaction, \emph{``Once I taught it the process, next time I only needed to provide a few keywords to complete the task.''} P1 found the context supplement very helpful, which allows for queries without the need for \emph{``careful consideration''}. However, P1 also found the dialogues somewhat verbose, stating, \emph{``Occasionally, the agent would ask me for unnecessary parameter details. For example, asking me for the WeChat version number—I don't even know it. I would prefer the agent to generate some reliable options for me to choose from.''}} 

\textcolor{minorrevise}{
All participants indicated that they preferred demonstration over instruction editing. P2 commented, \emph{``As long as the text isn't too far off, I won't modify it. Editing the text is less convenient than just demonstrating the operation later.''}}

\textcolor{minorrevise}{
Regarding demonstrations, participants (10/12) commonly noted that most interventions were for locating small icons, which they accepted. For improvement, P1 suggested incorporating images into the step descriptions for clarity, similar to the online tutorials including screenshots. Some participants found that the agent's prompts were sometimes incorrect, leaving them unsure about how to proceed. P8 mentioned, \emph{``The agent asked me to click `workshop', but there was nothing like that on the page.''} P4, an older user, remarked, \emph{``Adding steps is quite challenging for me; I hope the tutorials could be more accurate.''}}

\textcolor{minorrevise}{
In addition to the above interaction methods, P3 expressed a desire for more intuitive communication methods, \emph{``Guess what I am trying to do and help me out when I am stuck in a certain step.''}} 



\item 
\textcolor{minorrevise}{\textbf{Tolerance for Latency:}
 Participants' tolerance for latency varied depending on the individual, interaction state, and task. Most participants (10/12) acknowledged that some complex tasks understandably took longer. P5 noted, \emph{``Although the process took a bit of time, the real-time feedback made it feel like a learning experience, which was acceptably paced for me.''}
 However, some participants expected lower latency for simpler tasks during their first use of Prompt2Task. P1 indicated that while he could accept the execution time, he hoped for faster response in chatting. P2 felt frustrated when the agent was slow in performing tasks she already knew how to complete. P4 remarked, \emph{``It felt like I had to wait a bit for each step, which made me feel like I was just idling, even though it was only a few seconds. Perhaps some animations or interactive elements could help improve the experience.''} Fortunately, after using Prompt2Task a second time, all participants expressed satisfaction with the completion time.}

\item 
\textcolor{minorrevise}{\textbf{Evaluation of Intelligence and Expectations:} All participants acknowledged the system's intelligence. P3 specifically praised the feature that \emph{``aligns the smartphone's operational page with each step description''}. Similarly, P5 stated, \emph{``Before, I struggled with the tutorial's first step not matching my current page. Now it is a significant improvement.''}. Room for improvement was also noted. P9 advocated for a more \emph{``aggressive''} execution approach, arguing the current system is slightly conservative. P11 noted that some steps for confirmation were unnecessary. P12 observed, \emph{``Tasks generally succeeded, but some took a roundabout path.''}}

\item 
\textcolor{minorrevise}{\textbf{Application Scenarios:} Many participants praised the system's transparency and user-centered approach. P8 stated, \emph{``It gives me a feeling like I'm navigating through the tasks. It's very interesting.''} The utility of Prompt2Task in simplifying otherwise complex tasks was another strong point. P1 expressed, \emph{``I often ask my family how to use various functions on my phone, but it's challenging to follow their instructions. Now I can just copy their responses and the system executes the steps for me.''} Similarly, P12 shared, \emph{``I find new applications, especially those with complex functions, very daunting. It's such a hassle to contact customer service and then follow their steps. Prompt2Task saves me from all that trouble.''} P9 pointed out, \emph{``I can define a few routine tasks, like accounting after spending money. The agent can then automate these tasks for me.''} P3 stated, \emph{``I often forget the steps, so the replay is very useful for me.''}}

\end{itemize}


\section{Discussion}

In this section, we delve deeper into the significance of knowledge and the applications of our proposed Prompt2Task. We also discuss the limitations of the system and potential future work.

\subsection{Open and Accumulated Knowledge}

\label{sec:knowledgediscussion}

Prompt2Task aims to convert information across modalities (from texts to operations), with its stages focusing on bridging gaps between them and minimizing information loss. In addition to the collaboration of its multi-agents, Prompt2Task's effectiveness is derived from its open and accumulated knowledge, ensuring its broad applicability and efficiency.

Prompt2Task leverages massive online tutorials to address a broader range of user needs, surpassing existing work that relies on pre-collected datasets \cite{li_mapping_2020, venkatesh_ugif_2022, 10.1145/3581998}. It supports functionalities deep within applications, which exceeds the capabilities of current assistants like Google Assistant and HUAWEI Celia. Moreover, Prompt2Task's approach to mobile semantics is open-ended, enabling exploration beyond rigid adherence to given instructions. The open-ended approach allows Prompt2Task to handle diverse textual inputs and adapt to changing environments, significantly expanding the task automation's application scope.



Prompt2Task's knowledge base evolves through user usage and feedback. The evolution manifests in improved success rates, rising from 60.80\% to 95.24\%, and reduced completion time for repeated tasks (Figure \ref{fig:timemethod}). Notably, the accumulation of knowledge contributes to a reduction in user interventions. When similar task requirements arise again, Prompt2Task operates autonomously, only requiring user intervention for contextual parameter adjustments or significant interface updates. The accumulated knowledge empowers Prompt2Task to tackle increasingly complex and nuanced tasks, dynamically adapting to emerging user needs.



\begin{sloppypar}
\textcolor{revise}{It is natural to consider how Prompt2Task handles outdated knowledge. While accumulated knowledge enhances Prompt2Task's performance by improving efficiency (Cases 1 and 4 in Appendix \ref{apd:cases}), reducing user interaction in future tasks (Cases 2 and 3 in Appendix \ref{apd:cases}), and increasing success rates (Case 5 in Appendix \ref{apd:cases}), it is not indispensable. When a significant app update causes the accumulated knowledge outdated, Prompt2Task's multi-agent design can still work effectively. As shown in Figures \ref{fig:informationcollection}, \ref{fig:instructiongeneration}, and \ref{fig:operationmapping}, outdated or missing knowledge only affects specific parts. Therefore, the agents still have the potential to make correct decisions. Upon successful execution, the accumulated knowledge is updated.}
\end{sloppypar}

\begin{table}[t]
    \caption{\textcolor{revise}{``Outdated tutorials'' is the result of using the 500 ground-truth step descriptions (100 step descriptions over 5 rounds of testing) as input. ``Tutorials not found'' is the evaluation of the entire pipeline, which used a total of 12,500 textual prompts (2,500 prompts over 5 rounds of testing) as input.}}
    \label{tab:outdatedknowledge}
    \centering
    \begin{tabular}{lcccc}
		\toprule  
            &\multicolumn{2}{c}{\textbf{Outdated Tutorials}}&\multicolumn{2}{c}{\textbf{Tutorials Not Found}}\\
		&Success Rate&User Intervention&Success Rate&User Intervention\\
    \midrule
         Prompt2Task&75/155&-&611/1448&- \\
         Prompt2Task(AK)&-&-&467/967&- \\
         Prompt2Task(minimal UI+AK)&140/155&0.75(sd = 1.08)&698/855&0.68(sd = 1.22) \\
         Prompt2Task(UI+AK)&140/155&0.75(sd = 1.08)&91/110&0.86(sd = 1.21) \\
    \bottomrule
    \end{tabular}

\end{table}

\textcolor{revise}{As illustrated in Table \ref{tab:outdatedknowledge}, Prompt2Task demonstrates high success rates even with outdated tutorials or when no tutorials are found, indicating that the \emph{Grounding Agent} does not entirely rely on tutorials. In the worst-case scenario, where no tutorials match, no useful accumulated knowledge is available, and no human intervention is provided, Prompt2Task still achieves a 42.20\% success rate. Even if the system fails, stage-wise transparency allows users to help the agent complete the task correctly.}


\subsection{Applications}




\begin{sloppypar}
By incorporating massive external knowledge and accepting various textual inputs, Prompt2Task supports a wide range of task executions, benefiting both end users and developers.
\end{sloppypar}


For end users, Prompt2Task improves the usability of smartphones and facilitates assistance between peers. It offers guidance for users to navigate unfamiliar smartphone functions, enabling them to access and automate tasks deep within applications. It allows users to personalize their smartphone experience by automating their own complex or repetitive tasks. Furthermore, textual assistance from various sources, such as family, friends, or online customer service, can be fed to Prompt2Task for automating tasks, which is valuable for teamwork or external support. For instance, an elderly user receiving textual assistance from family members can simply input the text into Prompt2Task for automating tasks, without the need to repeatedly switch between text and complex interfaces.






For developers, Prompt2Task can reduce the costs of developing and maintaining task automation. It can cost-effectively transform a vast array of existing tutorial texts into operations, seamlessly integrated into current voice assistants or online customer service, thereby expanding the range of supported tasks. Application and platform developers can leverage Prompt2Task to transform text-based cases in bulk into interactive tutorials or executable test scripts. It significantly reduces the cost of updating voice assistants, interactive tutorials, and test scripts, as developers need only update the source texts regularly to respond to the latest application functions and interfaces. Nonetheless, in sensitive applications such as banking, implementing robust authentication mechanisms is crucial to prevent malicious command execution.

Additionally, although Prompt2Task is designed for smartphones, its underlying methodology, particularly the multi-agent framework and extensive knowledge base, can be adapted to automate tasks in other domains, such as websites and IoT devices.



\subsection{Limitations and Future Work}

We outline the limitations of Prompt2Task, along with possible directions for future enhancements.

\textbf{Adapting Dialogue Mechanisms for Diverse User Groups:}
The current design of the \emph{Analysis Agent} offers a generalized approach without catering to the specific needs of different user demographics. For instance, older and younger users may express issues differently, necessitating a more nuanced, tailored dialogue system to accommodate their needs.

\textbf{Parsing Complex Tutorial Logic:}
Automating responses to conditional branches like ``If the version number is less than 14.0, please upgrade the app'' could enhance the system's efficiency. Future work will focus on more sophisticated logic analysis, including nested conditions.

\textbf{Improving Semantic Understanding of GUI:}
In the present design, the mobile semantics module emphasizes the textual descriptions of mobile pages. However, current models often fall short, particularly when faced with pages not meeting accessibility standards. \textcolor{revise}{It results in blocking access to operation widgets and a notable loss of information during the translation from interface visuals to text.} Moving forward, we aim to enhance algorithms that better recognize semantic structures within GUI, with the potential improvement of image recognition technologies.


\textbf{Balancing Breadth with Depth:} Currently, to expand the application scope, Prompt2Task applies the capabilities of LLM across various stages. However, some domains have not been deeply explored, \textcolor{revise}{such as addressing ambiguities in natural language (which the LLM's rephrasing might not address)} and setting instruction parsing templates to reduce the risk of LLM fabrication. As Prompt2Task's usage data increases, we will systematically collect domain-specific data and user feedback to fine-tune the LLM or train local models, gradually reducing reliance on generalized templates and increasing its capacity for nuanced, domain-specific processing.


\section{Conclusion}
This paper presents Prompt2Task as an effective solution in UI
 task automation, particularly addressing the barriers of technical complexity, thereby facilitating its broader application. Utilizing the multi-agent framework and supplementing online tutorials, Prompt2Task is capable of interpreting unrestricted textual prompts to perform a wide range of automation tasks. As knowledge is accumulated from user usage and feedback, Prompt2Task improves task success rates and efficiency. Performance evaluations showcase a marked rise in success rates, ascending from an initial 22.28\% to a remarkable 95.24\%, with an average of only 0.69 user interventions per new task. Even without user interventions, a success rate of 78.16\% is achieved. User studies further validate the effectiveness of Prompt2Task in delivering a user-friendly and intelligent experience. Upcoming efforts aim to minimize the burden of user interaction, engage a wider user group, and boost the system's proficiency in managing complex tasks.

\begin{acks}
This work is supported by the Natural Science Foundation of China under Grant No. 62132010, Beijing Key Lab of Networked Multimedia, Institute for Artificial Intelligence, Tsinghua University (THUAI), Beijing National Research Center for Information Science and Technology (BNRist), 2025 Key Technological Innovation Program of Ningbo City under Grant No. 2022Z080, Beijing Municipal Science and Technology Commission, Administrative Commission of Zhongguancun Science Park No.Z221100006722018, and Science and Technology Innovation Key R\&D Program of Chongqing.
\end{acks}
 
\bibliographystyle{ACM-Reference-Format}
\bibliography{sample-base}

\appendix
\section{Mult-Agent Design}

The following provides the technical and prompt details for each agent.

\label{apd:multiagent}
\subsection{Analysis Agent}
\label{apd:analysisagent}
\begin{figure}[htbp]
\centering
    \includegraphics[width=\textwidth]{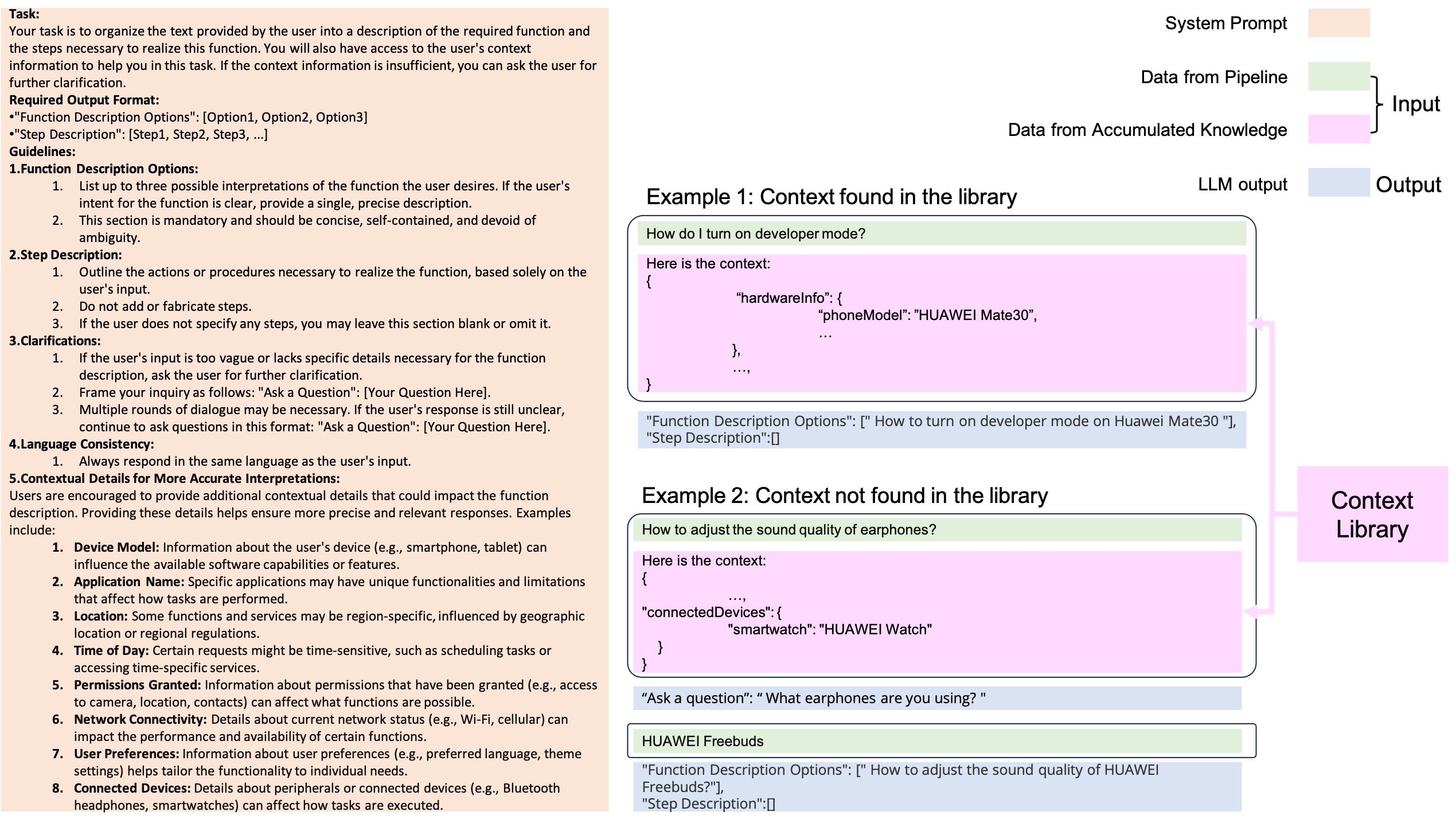}
  \caption{\textcolor{revise}{Prompts used by the \emph{Analysis Agent}.
  }}
  \label{fig:analysisprompt}
\end{figure}

\textcolor{revise}{The agent processes the input through GPT-4 to produce a function description and step description as reflected in the input, with prompt details shown in Figure \ref{fig:analysisprompt}.}

\subsection{Retrieval Agent}
\begin{figure}[htbp]
\centering
    \includegraphics[width=\textwidth]{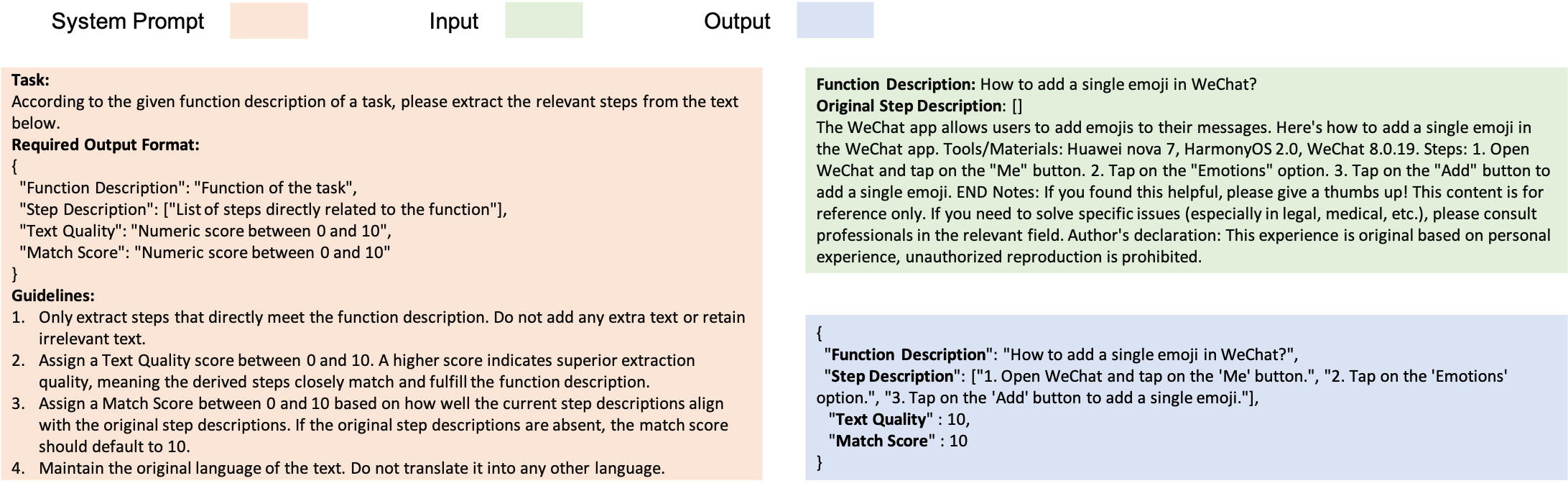}
  \caption{\textcolor{revise}{Prompts used by the \emph{Retrieval Agent}.
  }}
  \label{fig:retrieveprompt}
\end{figure}

\textcolor{revise}{The agent uses GPT-4 to extract the step description from the raw text of search results, with prompt details shown in Figure \ref{fig:retrieveprompt}.}

\subsection{Parsing Agent}
\begin{figure}[htbp]
\centering
    \includegraphics[width=\textwidth]{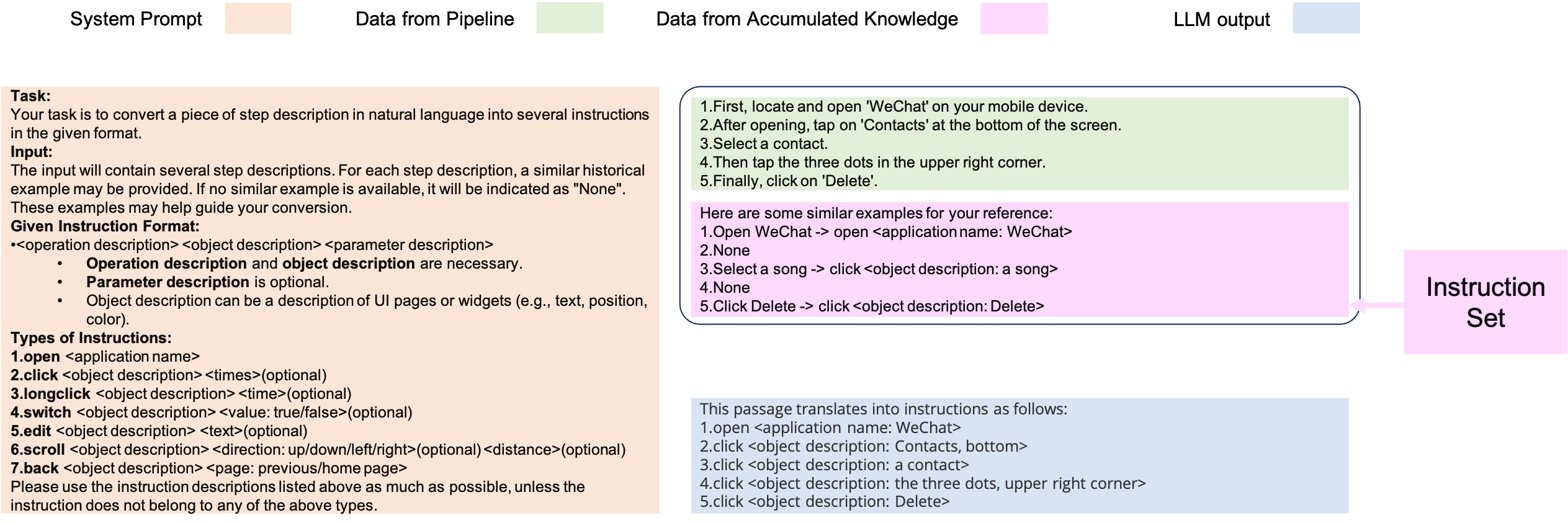}
  \caption{\textcolor{revise}{Prompts used by the \emph{Parsing Agent}.
  }}
  \label{fig:parsingprompt}
\end{figure}

\textcolor{revise}{The agent uses GPT-4 to accomplish the transformation from step descriptions in natural language to formalized instructions, with prompt details shown in Figure \ref{fig:parsingprompt}.}

\subsection{Grounding Agent}
\label{apd:groundingagent}
\textcolor{revise}{In this section, we provide detailed descriptions of the strategies employed by the \emph{Grounding Agent}. Initially, based on the current page's layout hierarchy, the agent identifies and selects all widgets that either possess text descriptions or have interactive attributes, thereby constructing a set of potential operation candidates. Subsequently, the agent employs the following strategies to predict the operational widget within these candidates:}

\begin{figure}[htbp]
\centering
    \includegraphics[width=\textwidth]{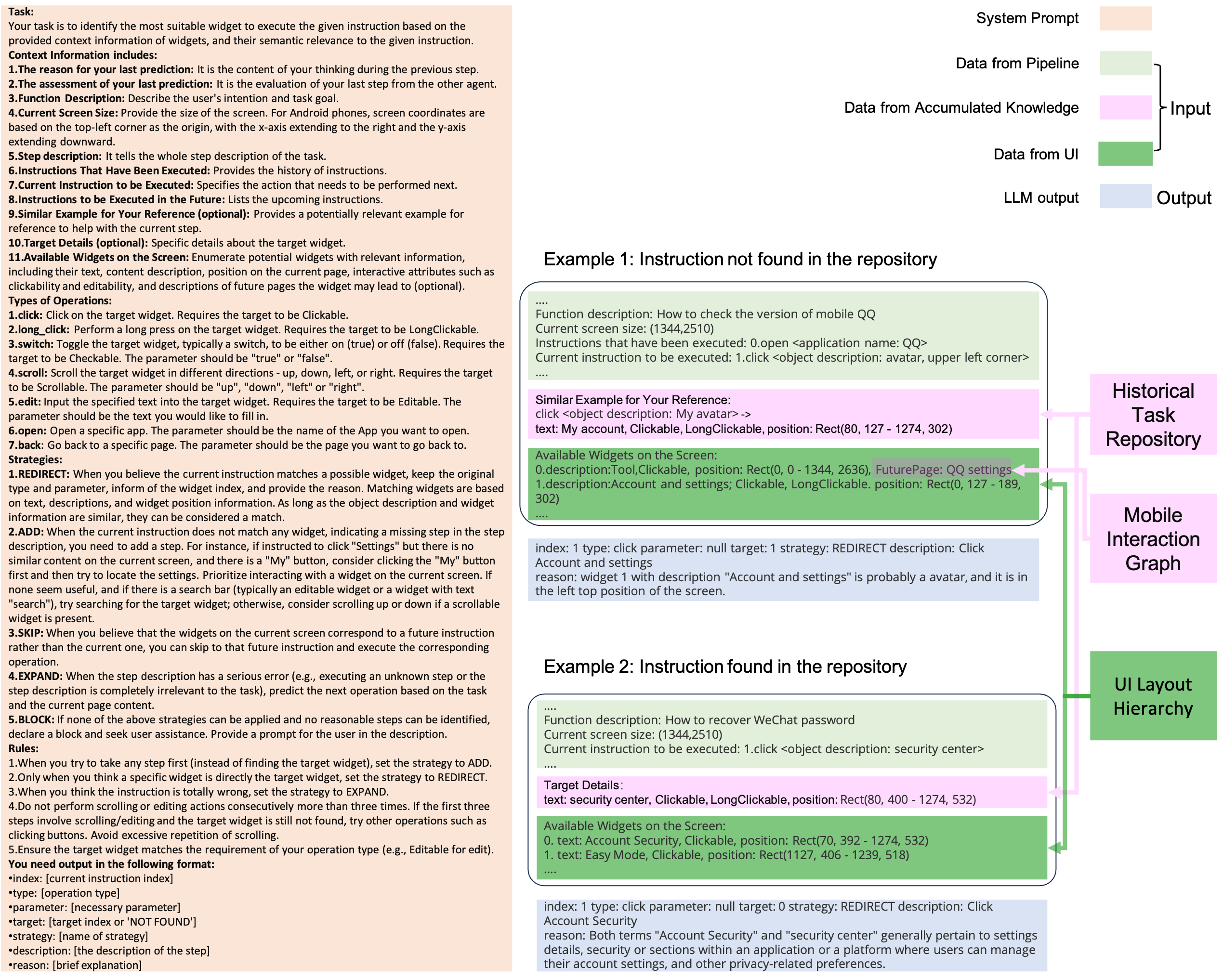}
  \caption{\textcolor{revise}{Prompts used by the \emph{Grounding Agent}.
  }}
  \label{fig:groundingprompt}
\end{figure}

\textcolor{revise}{\begin{enumerate}
    \item \textbf{Historical task invocation}: If the instruction has been previously mapped to an operation, the agent can refer to stored widget data in the historical task repository. It enables the agent to direct the operation to a corresponding widget on the current interface. To evaluate the similarity between the candidate widget and the stored widget, we employ a heuristic rule-based similarity function, as described in the work of \citeauthor{10.1145/3610929} \cite{10.1145/3610929}. This function considers 26 features of a widget, including its style and position in images, its style and position in layouts, and its role in page segmentation. The similarity thresholds have been established using Receiver Operating Characteristic (ROC) curve analysis.
    \item \textbf{New instruction-based grounding}: For new instructions, the agent then employs the following strategies:
    \begin{enumerate}
        \item \textbf{Strict text matching}: It involves searching for a widget whose text closely matches the object description ($\text{Instruction}.\text{Object}$). Similarity is defined as Equation \ref{eq:textsimilarity}, where $\text{Widget}.\text{text}$ denotes the text and content description contained in the widget's layout information, and $LCS$ signifies the length of the longest common sub-string between two strings. A match is confirmed only if the similarity exceeds an empirically preset threshold.
        \begin{equation}
        \label{eq:textsimilarity}
        \begin{aligned}
        Sim(\text{Widget}, \text{Instruction}.\text{Object}) &= \frac{2 \times LCS(\text{Widget}.\text{text}, \text{Instruction}.\text{Object}.\text{text})}{length(\text{Widget}.\text{text})+length(\text{Instruction}.\text{Object}.\text{text})}
        \end{aligned}
        \end{equation} 
        \item \textbf{Redirect, Add, Skip, Expand, Block}: The agent enumerates potential widgets with relevant information, including their text, content description, position on the current page, interactive attributes and so on. Subsequently, the LLM considers the current context to choose a strategy and then predicts the most suitable widget for executing the operation. Details of the prompt are illustrated in Figure \ref{fig:groundingprompt}.
    \end{enumerate}
\end{enumerate}}


\subsection{Assessment Agent}
\label{apd:assessmentagent}

\begin{figure}[htbp]
\centering
    \includegraphics[width=\textwidth]{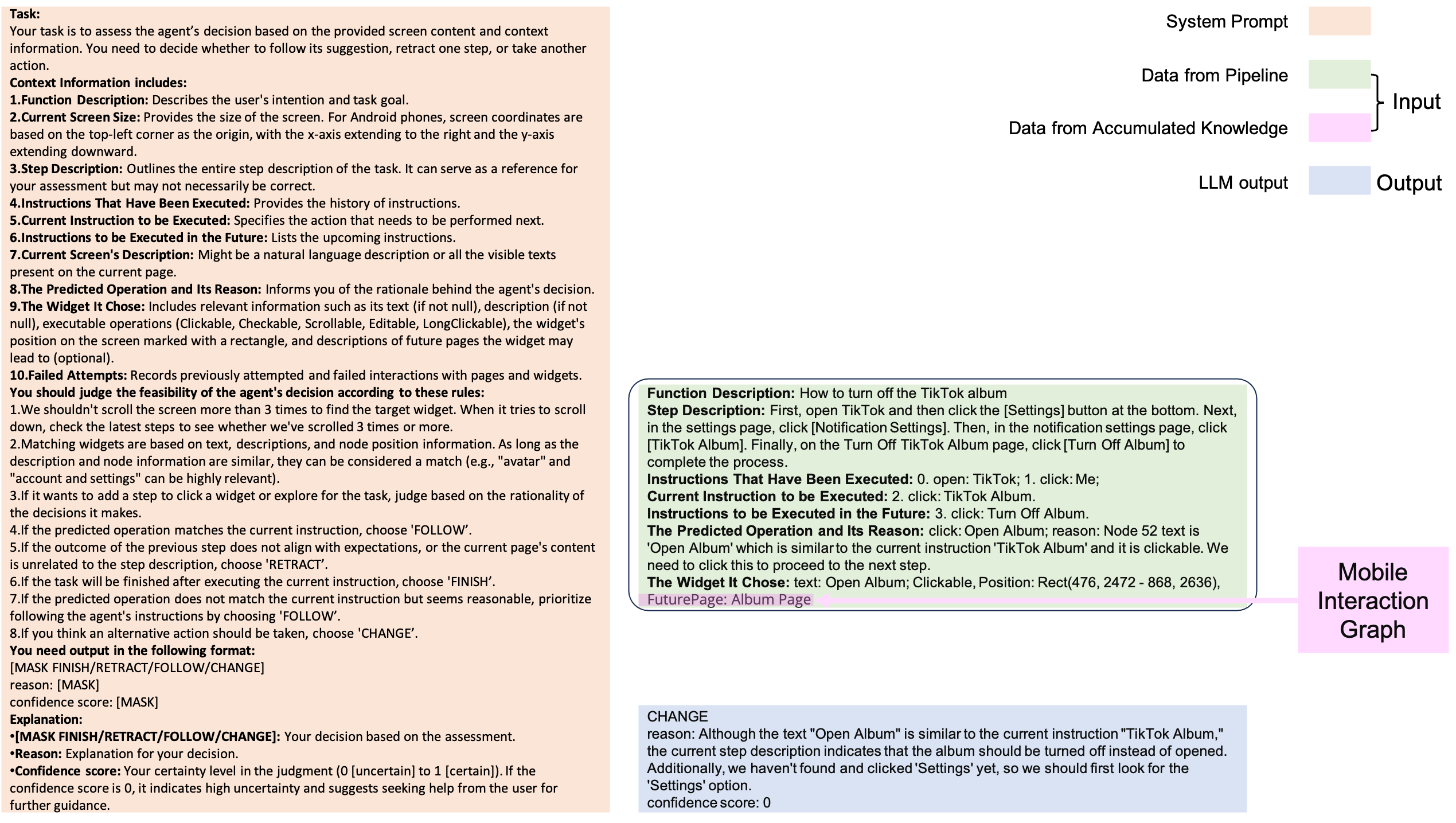}
  \caption{\textcolor{revise}{Prompts used by the \emph{Assessment Agent}.
  }}
  \label{fig:assessmentprompt}
\end{figure}

\textcolor{revise}{The \emph{Assessment Agent} utilizes the LLM with a rich set of contextual information to make decisions: follow, change, retract, or finish, while also describing if it is confident for its decision, as shown in Figure \ref{fig:assessmentprompt}. If the agent is unconfident in its decision, the system automatically prompts the user for assistance.}

\section{Accumulated Knowledge}

\subsection{Construction of the Historical Task Repository}
\label{apd:historicalrpa}


The Historical Task Repository serves as an archive for successfully executed automation tasks. Since the \emph{Grounding Agent} doesn't strictly adhere to pre-generated instructions, the repository focuses on storing the actual sequences of operations executed. This subsection details how these operation are reverse-engineered into corresponding instructions and step descriptions.

\begin{enumerate}
    \item \textbf{Operation Sequence Storage}: The first step is
    to archive the actual sequence of operations executed by the \emph{Grounding Agent}, which is the most reliable record of the automation task. The system archives all layout hierarchical information and screenshots of the pages visited. It also logs the operation object's index in the page's layout tree, which can be used to retrieve the object's layout information. Additionally, since the object's position is recorded, it allows for capturing its corresponding image on the full-page screenshot.
    
    \item \textbf{Reverse Engineering to Instructions}: Using the stored operation sequence, the system reverse-engineers to produce the corresponding instructions. Each operation is mapped to the potential instruction that could have prompted it, adhering to the format specified in Table \ref{tab:operation}. In this context, the operation types and parameters have already been logged by the system during the triggering of the operation. The descriptions of operation objects are derived from their texts, content descriptions, positions, and the average RGB values of their corresponding images.
    
    \item \textbf{Generation of Step Descriptions}: Alongside instructions, the system also formulates step descriptions. They are rephrased versions of the instructions, articulated in natural language, generated by GPT-4.
\end{enumerate}

\subsection{Construction of the Context Library}
\label{apd:contextlibrary}

The context library plays a crucial role in enhancing the system's capacity to interpret vague textual prompts by providing an array of mobile-specific parameters. Here is the method employed for constructing the library.

\begin{enumerate}
    \item \textbf{Online Q\&A Data Collection}: 
    Initially, we sourced a set of 20,265 questions related to mobile phone usage from a recognized online Q\&A platform\footnote{\href{https://zhidao.baidu.com/}{https://zhidao.baidu.com/}}. 
    
    \item \textbf{Participant Involvement}:
    The collected questions were distributed among 24 undergraduate college students who participated in a formative study.
    
    \item \textbf{Contextual Parameter Extraction}:
    Each participant was tasked with identifying the ``central'' keywords in each question. They then examined various variations of the question to infer the contextual parameters required for comprehensive understanding.
    
    \item \textbf{Context Library Formation}:
    We employed experts to review, analyze, and categorize the list of contextual parameters identified by the participants. Their insights were integrated to develop the finalized context library.
\end{enumerate}

\begin{figure}[htbp]
\centering
    \includegraphics[width=\textwidth]{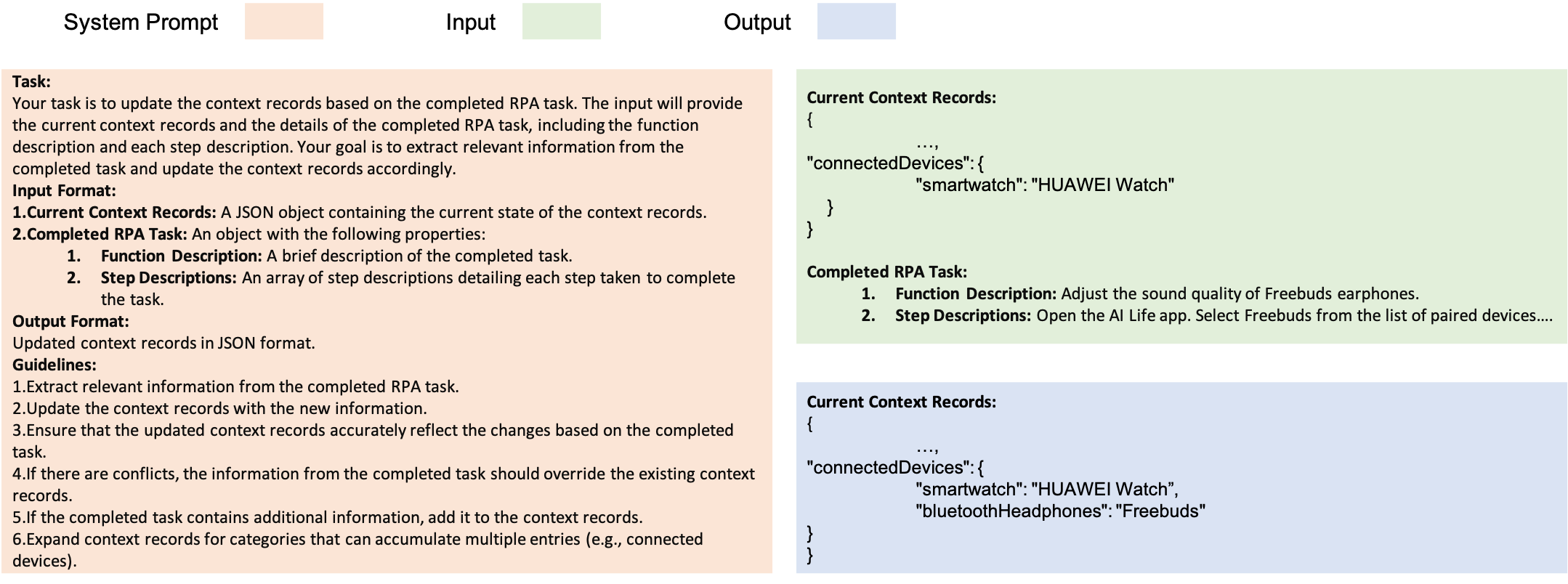}
  \caption{\textcolor{revise}{Prompts used to update the context library based on the currently completed automation task.}}
  \label{fig:contextprompt}
\end{figure}

\textcolor{revise}{Here is an example of records in the context library. Each time a task is successfully completed, the context library integrates new information using GPT-4, with prompt detailed in Figure \ref{fig:contextprompt}.}
\begin{verbatim}
{
    "hardwareInfo": {
        "deviceName": "mySmartPhone",
        "phoneModel": "HUAWEI Mate30",
        "battery": "0.23",
        "availableWifi": ["Network1", "Network2"]
    },
    "systemSetting": {
        "display": {
            "darkmode": true,
            "autoBrightness": true
        }
    },
    "privacyAndSecurity": {
        "locationAllowedServices": {
            "allowed": ["Maps", "WeChat", "Lark"]
        },
        "permissions": {
            "camera": true,
            "location": true,
            "contacts": false
        }
    },
    "usingApp": ["App1", "App2"],
    "operatingSystem": "HarmonyOS 2.0",
    "location": "Beijing, China",
    "timeOfDay": "2023-04-12T14:30:00Z",
    "networkConnectivity": "Wi-Fi",
    "userPreferences": {
        "preferredLanguage": "en",
        "themeSettings": "dark"
    },
    "connectedDevices": {
        "bluetoothHeadphones": "HUAWEI Freebuds",
        "smartwatch": "HUAWEI Watch"
    }
} 
\end{verbatim}

\subsection{\textcolor{revise}{Typical Cases}}
\label{apd:cases}

\textcolor{revise}{This section illustrates typical cases where accumulated knowledge optimizes the system in various ways, while also explaining how the system handles outdated knowledge.}


\subsubsection{\textcolor{revise}{Case 1}}

\textcolor{revise}{Case 1 demonstrates the optimization of the \emph{Grounding Agent} by the historical task repository.}

\begin{figure}[htbp]
    \centering
    \includegraphics[width=.5\linewidth]{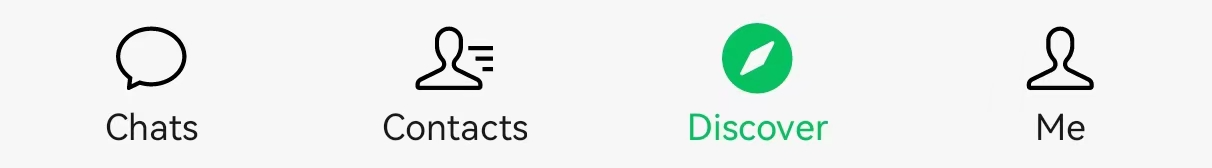}
      \caption{\textcolor{revise}{Case 1: The instruction indicates clicking ``Account'', but the current UI suggests that ``Me'' should be clicked instead.}}
      \label{fig:case1}
\end{figure}

\textcolor{revise}{Suppose the current instruction is <type: click, object: ``Account''>. As shown in Figure \ref{fig:case1}, the current UI suggests that ``Me'' should be clicked based on semantic similarity. During the initial execution, the \emph{Grounding Agent} uses the LLM and chooses the redirect strategy to locate the target widget. This process is relatively slow and may result in inconsistent outcomes. If the target widget cannot be found, the \emph{Grounding Agent} will choose to block, prompting the user to demonstrate the correct operation.}

\textcolor{revise}{After the first execution, in subsequent executions, the system leverages the Prompt2Task model stored in the historical task repository. Knowing that this step is clicking ``Me'', the \emph{Grounding Agent} can directly locate the target widget using heuristic rules, resulting in faster and more reliable outcomes.}

\textcolor{revise}{If the app undergoes significant updates such that ``Me'' no longer exists or the widget style changes drastically, making it difficult to locate the widget using heuristic rules, the \emph{Grounding Agent} will refer to the instruction <type: click, object: ``Me''> and use the LLM to infer the operation again (using strategies such as redirect, add, skip, expand, or block). Even if the knowledge in the historical task repository is outdated, it still provides guidance by including more target details in the prompt, as shown in the prompt engineering of the \emph{Grounding Agent} in Appendix \ref{apd:groundingagent}. After completing the new task, the historical task repository will store the updated target details.}





\subsubsection{\textcolor{revise}{Case 2}}


\textcolor{revise}{Case 2 demonstrates the optimization of the mobile semantics module in the \emph{Grounding Agent} by the historical task repository, which then enhances the agent's prediction.}

\begin{figure}[htbp]
    \centering
    \includegraphics[width=.52\linewidth]{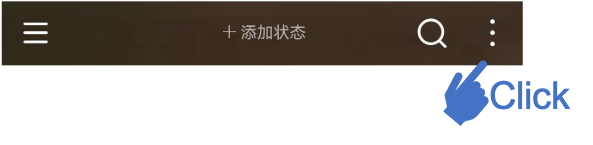}
      \caption{\textcolor{revise}{Case 2: The instruction indicates clicking ``more'', but the icon's information is insufficient for the agent to locate it, likely requiring user demonstration upon first encounter.}}
      \label{fig:case2}
\end{figure}

\textcolor{revise}{Suppose the current instruction is <type: click, object: ``more''>, and the current UI is as shown in Figure \ref{fig:case2}. During the initial execution, the agent is unable to identify ``more'' due to insufficient UI information (no text labels and pixel-based recognition failed). Therefore, the \emph{Grounding Agent} enters a block state and requests the user to demonstrate the correct operation.}

\textcolor{revise}{After the first execution, in subsequent executions, the mobile semantics module learns from the Prompt2Task model stored in the historical task repository and knows that the icon image corresponds to the ``more'' label. Therefore, it can provide this information to the \emph{Grounding Agent}, which can then use strict text matching to directly locate and click the widget.}

\textcolor{revise}{As long as the icon image does not change significantly, the mobile semantics module can label the icon as ``more''. Otherwise, the process of requiring user intervention to identify the icon will be repeated, and after task completion, the historical task repository will be updated accordingly.}




\subsubsection{\textcolor{revise}{Case 3}}

\textcolor{revise}{Case 3 demonstrates the optimization of the \emph{Analysis Agent} by the context library.}

\textbf{
\begin{figure}[htbp]
    \centering
    \includegraphics[width=\linewidth]{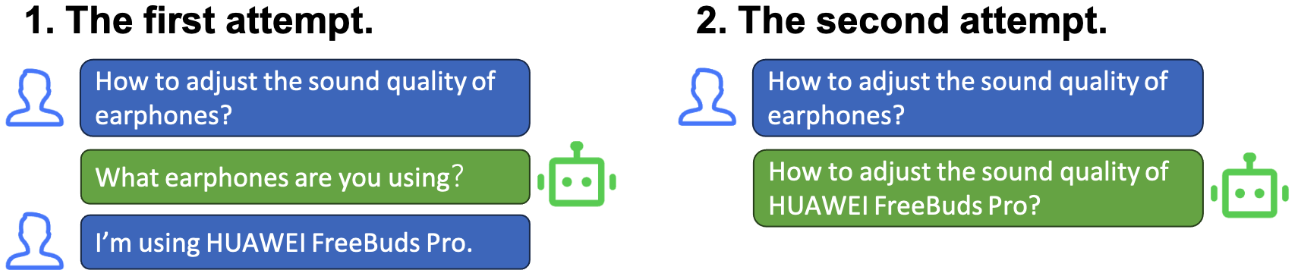}
      \caption{\textcolor{revise}{Case 3: An example of auto-completing the textual prompt.}}
      \label{fig:case3}
\end{figure}}

\textcolor{revise}{As shown in Figure \ref{fig:case3}, the user input is ``How to adjust the sound quality of earphones''. During the initial encounter, the \emph{Analysis Agent} needs to query the user for the earphones' model. Once the user provides the information, the context library integrates it, as described in Appendix \ref{apd:contextlibrary}.}

\textcolor{revise}{In subsequent instances of similar tasks, the agent will proactively ask the user if he/she needs to ``adjust the sound quality of HUAWEI Freebuds Pro'', requiring only a simple confirmation from the user.}

\textcolor{revise}{If the user's context changes, such as a different earphone model, the context library may provide incorrect information. In such cases, the user can reject the agent's recommendation, and the agent will engage in a dialogue with the user to correct and then update the context library.}





\subsubsection{\textcolor{revise}{Case 4}}

\textcolor{revise}{Case 4 demonstrates the optimization of the \emph{Parsing Agent} by the instruction set.}

\begin{figure}[htbp]
	\centering
	\subfigure[]{
		\begin{minipage}[b]{0.48\textwidth}
            \label{fig:case4_1}
		\centering
			\includegraphics[height=4cm, keepaspectratio]{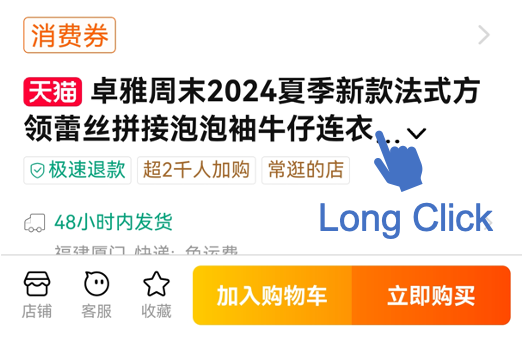}
		\end{minipage}
	}
    \subfigure[]{
            \begin{minipage}[b]{0.48\textwidth}
            \label{fig:case4_2}
		\centering
		\includegraphics[height=4cm, keepaspectratio]{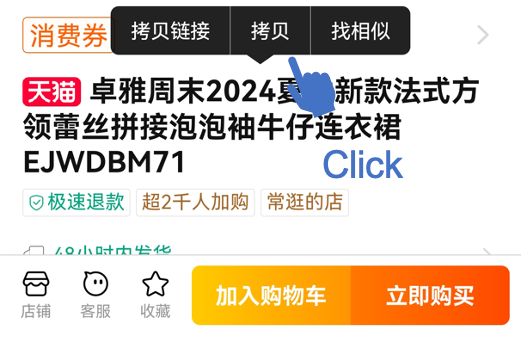}
              \end{minipage}
    }
        \caption{\textcolor{revise}{Case 4: The instruction is a complex copy command that requires first long-pressing the title and then selecting ``copy'' from the options that pop up.}}
\end{figure}

\textcolor{revise}{Suppose the current step description is ``copy the title'', which is a complex instruction that should be decomposed into two instructions:
\begin{itemize}
\item <type: longclick, object: title widget> (Figure \ref{fig:case4_1})
\item <type: click, object: ``copy''> (Figure \ref{fig:case4_2})
\end{itemize}}


\textcolor{revise}{During the first execution, since it is not included in the basic instruction definition, the \emph{Parsing Agent} can only generate <type: copy, object: ``title''>. Consequently, the \emph{Grounding Agent} cannot match the corresponding operation and must rely on the ``add'' strategy or user demonstration to successfully execute the task. After task completion, the user maps the original compound instruction to the generated decomposed instructions, and the instruction set stores this correspondence for future use.}

\textcolor{revise}{In subsequent executions, the \emph{Parsing Agent} will automatically decompose the compound instruction by finding similar examples in the instruction set. Then it will allow the \emph{Grounding Agent} to potentially find the target widget directly through heuristic rules. Even if it cannot, it will reduce the difficulty of LLM reasoning in the \emph{Grounding Agent}.}

\textcolor{revise}{If the app undergoes significant updates, changing the execution steps of this compound instruction, the instructions generated by the \emph{Parsing Agent} may be incorrect. Consequently, the \emph{Grounding Agent} will rely on the LLM to redeploy strategies (redirect, add, skip, expand). If it encounters a block, it will require user demonstration.}




\subsubsection{\textcolor{revise}{Case 5}}

\textcolor{revise}{Case 5 demonstrates the optimization of the \emph{Grounding Agent} and the \emph{Assessment} Agent by the mobile interaction graph.}

\begin{figure}[htbp]
    \centering
    \includegraphics[width=.5\linewidth]{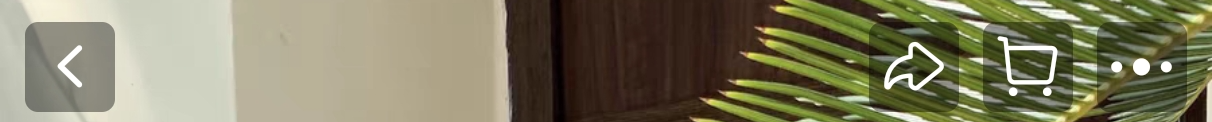}
      \caption{\textcolor{revise}{Case 5: The current UI contains 4 candidate widgets. The agents will consider various information for each widget, including the triggered future pages, to make the final decision.}}
      \label{fig:case5}
\end{figure}

\textcolor{revise}{Suppose the current step description is ``share the link with a friend'', and the current UI is shown in Figure \ref{fig:case5}. There are four interactive widgets available for operation from left to right. Suppose the mobile interaction graph already contains the resulting pages after triggering these widgets (with the mobile semantics module in the \emph{Grounding Agent} generating textual descriptions for each page):}

\textcolor{revise}{\begin{itemize}
    \item 0 $\rightarrow$ search result page
    \item 1 $\rightarrow$ forwarding list
    \item 2 $\rightarrow$ cart popup
    \item 3 $\rightarrow$ tools page
\end{itemize}}

\textcolor{revise}{As shown in Figure \ref{fig:groundingprompt} and Figure \ref{fig:assessmentprompt}, this information is incorporated into the prompts for the \emph{Grounding Agent} and the \emph{Assessment Agent}, aiding them in making accurate predictions and judgments.}

\textcolor{revise}{If a significant app update alters the page transition relationships, the mobile interaction graph might provide incorrect information, potentially leading both agents to erroneous decisions. If multiple erroneous attempts occur, the \emph{Assessment Agent} will request user intervention.}



\section{User Evaluation}
\subsection{Objective Results}
The impact of methods and participant groups on task completion time is shown in Table \ref{tab:timemethod}.

\begin{table}[htbp]
	\centering
	\caption{Impact of different methods and participant groups on task completion time (in seconds).}
        \label{tab:timemethod}
	\begin{tabular}{cccc}
		\toprule   
		&Manual& Prompt2Task (First Time) & Prompt2Task (Second Time)\\
		\midrule  
		Unskilled&110.47 (sd = 54.86)& 249.45 (sd = 84.47) & 33.11 (sd = 6.14)\\
            Skilled&88.86 (sd = 55.51)& 212.09 (sd = 83.66) & 32.49 (sd = 7.83)\\
		\bottomrule  
	\end{tabular}
\end{table}

\subsection{Subjective Feedback}

After the participants completed the experiment, we conducted interviews with them. The results of the 7-point Likert scale are shown in Table \ref{tab:statement}.

\begin{table}[!h]
    \centering
    \caption{Subjective feedback.}
    \begin{tabular}{ll}
    
		\toprule  
        Statements &  Results \\
        \midrule
        Providing feedback to the system is effortless.& 5.42 (sd=1.16)\\
         You can finish the task easily.& 6.33 (sd=0.65)\\
         You find the time taken to be acceptable.& 5.83 (sd=1.47)\\
         The system's decisions are sufficiently intelligent. & 6.50 (sd=0.52)\\
         The system's requests for user assistance are appropriately made. & 6.08 (sd=0.51)\\
         You can feel free to query.& 6.67 (sd=0.65)\\
         The experience is good.& 6.25 (sd=1.06)\\
         You are willing to use our Prompt2Task.& 6.58 (sd=0.67)\\
         \bottomrule
    \end{tabular}
    \label{tab:statement}
\end{table}

\end{document}